\documentclass[12pt,english]{article}
\usepackage[T1]{fontenc}
\usepackage[utf8]{inputenc}
\usepackage{geometry}
\usepackage{amsmath}
\usepackage{amssymb}
\usepackage{pgfplots}
\usepgfplotslibrary{groupplots}
\usepackage{natbib}
\usepackage{tikz}
\newcommand{\indep}{\perp \!\!\!\!\perp}

\makeatletter
 
\DeclareMathOperator*{\argmin}{arg\,min} 
\usepackage{epsf}

\newtheorem{lemma}{Lemma}

\newtheorem{assumption}{Assumption}
\newtheorem{proposition}{Proposition}

\usepackage{babel}

\usepackage{setspace}
\makeatother

\usepackage{babel}
\begin{document}

\title{True and Pseudo-True Parameters}
\author{Isaiah Andrews, Harvey Barnhard, Jacob Carlson\thanks{Email: iandrews@mit.edu, hbarnhard@g.harvard.edu, jacob\_carlson@g.harvard.edu.  We thank Jiafeng Chen, Bas Sanders, Jesse Shapiro, Davide Viviano, participants in the SETA 2023 ET Lecture, participants in the Oxford 2024 Hicks Lecture, Peter Phillips, and three anonymous referees for helpful comments.}}

\maketitle

\abstract{Parameter estimates in misspecified models converge to pseudo-true parameter values, which minimize a population objective function.
Pseudo-true values often differ from quantities of economic interest, raising questions of how, if at all, they are relevant for decision-making.  
To study this question we consider Bayesian decision-makers facing a linear population minimum distance problem.  
Within a class of priors motivated by the minimum distance objective, we characterize prior sequences under which posteriors concentrate on the pseudo-true value. This convergence is fragile to small changes in priors, implying that pseudo-true values are relevant for decision-making only in special cases.  Constructive results are nevertheless possible in this setting, and we derive simple confidence intervals that guarantee correct average coverage for the true parameter under every prior in the class we study, with no bound on the magnitude of misspecification.\\
Keywords: Model Misspecification, Pseudo-True Values\\
JEL Codes: C10, C52}

\section{Introduction}

Empirical research in economics often begins by positing a model which relates quantities of economic interest to the distribution of observable data.  Researchers then use model-implied relationships, together with observed data, to construct estimates or bounds for parameters of interest.

Unfortunately, commonly-used models impose assumptions which are difficult to validate, and which are sometimes rejected outright.  For instance, some models impose functional form restrictions such as linearity, or distributional restrictions on latent error terms.  Others impose homogeneity across economic agents, or behavioral assumptions such as utility or profit maximization.  Finally, methods that aim to uncover causal or structural relationships impose assumptions regarding unconfoundedness of treatment and the scope for spillovers across units.  When we have reason to doubt these assumptions, it can be unclear how to interpret model-implied estimates or bounds.  Indeed, absent some restriction on model misspecification, the quantities of interest are necessarily unidentified, and we can learn nothing from data.

There is a large literature on model misspecification spanning statistics and economics. An influential econometrics literature, including \citet{white_maximum_1982}, \citet{hall_large_2003}, \citet{muller_risk_2013}, \citet{hansen_inference_2021}, \citet{andrews_misspecified_2023}, and \cite{kleibergen_double_2025}, studies the problem of inference under model misspecification, and avoids identification problems for the quantity of economic interest $\theta$ by shifting the focus to ``pseudo-true'' parameter values, defined as the minimizers of a population objective function.  Under mild conditions these papers show consistency of estimates for the pseudo-true value (in point-identified settings) or the pseudo-true identified set (in set-identified ones).  Moreover, these papers provide inference results, for example showing asymptotic normality of point estimates and deriving consistent standard errors for the pseudo-true value.  These results have been highly influential for empirical practice with, for instance, the ``sandwich'' standard error formula discussed by White (1982) now widely adopted in the context of maximum likelihood estimation.

While focusing on pseudo-true values allows us to provide statistical guarantees, it leaves open the question of how, if at all, these pseudo-true values relate to the original quantities of economic interest.  The literature studying inference on pseudo-true values discusses this tension, with \citet{white_maximum_1982} writing  ``[the estimator] converges to a well defined limit, and may or may not be consistent for particular parameters of interest.''  Similarly, \citet{muller_risk_2013} writes that ``[it] is important to keep in mind that the pseudo-true parameter of the misspecified model must remain the object of interest for ... inference to make sense'' and \citet{hansen_inference_2021} write that ``it is difficult to give economic interpretation to pseudo-true parameter values. Consequently, this limits interest in valid inference procedures for pseudo-true values.''

This paper revisits the distinction between true and pseudo-true parameter values.  To abstract from sampling uncertainty we consider a population minimum-distance problem in which the distribution of the data is perfectly observed.  We adopt a decision-theoretic, and specifically Bayesian, perspective to ask under what conditions the posterior distribution for $\theta$, given the distribution of the observable data, concentrates around the pseudo-true parameter.  Such concentration implies, under mild conditions, that 
Bayes decision rules converge to plug-in rules based on pseudo-true values.

We provide three main results.  First, we characterize a class of joint priors for the data distribution and $\theta$ such that the posterior density for $\theta$ is proportional to a transformation of the minimum distance objective function.  This proportionality implies that the minimum distance objective is sufficient for decision-making, and hence an optimal way to summarize the data.  This is a natural class of priors to consider in the context of minimum distance estimation, since it corresponds to a belief that the minimum distance objective captures all decision-relevant information.  We show under independence and conditional mean-zero restrictions on the implied prior for model misspecification, proportionality holds if and only if an implicit misspecification prior satisfies a rotation-invariance condition. 

Second, we characterize prior sequences in this rotation-invariant class such that the posterior distribution concentrates around the pseudo-true value.  These priors assume that the degree of misspecification is negligible, which seems implausible in many economic applications.  We further find that concentration is fragile, in the sense that seemingly small changes to the prior lead posterior concentration to fail dramatically.\footnote{While these priors imply that the degree of misspecification is negligible a-priori, our results cover cases where the data violate the model's over-identifying restrictions to an arbitrarily large extent.  Hence, posterior convergence to the pseudo-true value requires that the posterior continues to reflect that misspecification is as small as possible, even once we know that the model is badly wrong.}  When posterior concentration fails, researchers with different priors on the form and degree of misspecification will have different posteriors for $\theta.$

This naturally raises the question of whether it is possible to give positive results under the class of priors we consider. Our third main result constructs confidence intervals that guarantee correct coverage of $\theta$ under all priors satisfying our rotation-invariance condition.  Since we consider the population problem, there is no randomness due to sampling and we instead define coverage using the ex-ante probability under the prior.  Our misspecification-robust confidence intervals:
\begin{enumerate}
    \item are centered at the pseudo-true parameter value;
    \item require no researcher input beyond the choice of weighting matrix;
    \item have width proportional to the square root of a population $J$-statistic; and
    \item impose no ex-ante upper bound on the degree of misspecification.
\end{enumerate}
Since these intervals are valid under a class of priors motivated by the minimum distance objective, they are a natural option for summarizing misspecification-driven uncertainty in settings where researchers adopt a minimum distance approach.

Inference under model misspecification is closely related to the large literature on partial identification, and the practice of plugging in pseudo-true parameter estimates for decision-making is an instance of what \citet{manski_econometrics_2021} terms ``as-if optimization.''
In settings where we are concerned with model misspecification an alternative approach is to explicitly bound the possible degree of misspecification and derive results which are valid under all data generating processes satisfying this bound, for instance by characterizing the identified set for the quantity of interest.
We show, however, that computing identified sets under a bound on  the norm of the moments at the true parameter leads to counter-intuitive behavior in our setting, where the identified set narrows as the observed violation of the model's over-identifying restrictions grows worse, rather than widening, as our confidence intervals do.

The next section introduces our population minimum distance setting and formally defines model misspecification and pseudo-true values.  Section \ref{sec: decision rules} introduces the decision problem we study and provides our first main result, characterizing the class of priors such that the posterior for $\theta$ depends on the data through the minimum distance objective.  Section \ref{sec: covergence-inducing priors} characterizes sequences of priors in this class such that the posterior concentrates on the pseudo-true value, and shows that this concentration is fragile in important respects.  Finally, motivated by an invariance property derived in Section \ref{sec: decision rules}, Section \ref{sec: confidence sets} derives our suggested confidence intervals, compares them to identified sets based on bounds for the magnitude of misspecification, and considers several extensions to cases with sampling uncertainty.

\section {Setting}

\subsection{Minimum Distance Model}

Suppose that for a set $\mathcal{D}$ and $\Delta(\mathcal{D})$ the set of distributions on $\mathcal{D},$ a researcher observes a distribution $P\in\mathcal {P}\subseteq \Delta(\mathcal{D}).$  This corresponds to the large-sample limit of a setting where the researcher observes a sample of $n$ observations $D_{i}\in\mathcal{D}$ drawn iid from $P,$ since as $n\to\infty$ they can consistently estimate $P$ from $\{D_i\}_{i=1}^n.$  To abstract from sampling uncertainty  we consider the population problem where $P$ is directly observed, returning to the case with sampling uncertainty in Section \ref{sec: confidence sets}.\footnote{There are many empirical settings where the data may not be iid, or in which it is unclear that the data can reasonably be modeled as drawn from a fixed distribution \citep{phillips_laws_2003}. We abstract from these considerations in our analysis.} 

Further suppose that the researcher is interested in an economic quantity $\theta\in\mathbb{R}^p$, and has a model that implies that the true $(P,\theta)$ pair satisfies
\begin{equation}\label{eq: moment conditions}
g\left(\theta;P\right)=Y\left(P\right)-X\left(P\right)\theta=0 
\end{equation}
for known functions $Y:\mathcal{P}\to\mathbb{R}^{k}$ and $X:\mathcal{P}\to\mathbb{R}^{k\times p}.$ We assume that $X(P)$ has full column rank, and unless otherwise noted assume that the model is over-identified, with $k>p$.  We loosely refer to $g\left(\theta;P\right)$ as ``moments,'' though the linear minimum-distance setting we consider is more general than linear GMM. We focus on linear-in-parameters moments of the form \eqref{eq: moment conditions} for simplicity, but our exact results for this linear setting will translate to approximate results for models which can be linearly approximated, for instance under local misspecification as studied by \citet{armstrong_sensitivity_2021}. We again return to this connection in Section \ref{sec: confidence sets}.

\paragraph{Example: Linear IV} As a first example, suppose that $D_{i}=\left(Y_{i},X_{i},Z_{i}\right)$ for $Y_{i}\in\mathbb{R}$ a scalar outcome, $X_{i}\in\{0,1\}$ a binary endogenous treatment, and $Z_{i}\in\mathbb{R}^{k}$ a vector of $k$ mean-zero exogenous variables, $E[Z_i]=0$.  We assume that these data are generated from a potential outcomes model, where the potential outcomes $Y_i(x,z)$ may in general depend on both $X_i$ and $Z_i,$ and the potential treatments $X_i(z)$ may depend on $Z_i.$  The parameter of interest $\theta\in\mathbb{R}$ is the average treatment effect (ATE), 
\[
\theta=E[Y_i(1,Z_i)-Y_i(0,Z_i)],
\]
which captures the average effect on $Y_i$ from changing $X_i$ from zero to one.

If the researcher wants to estimate a constant-effect linear instrumental variables model with excluded instrument $Z_i$, this can be justified by assuming that $Z_i$ is excluded from $Y_i,$ $Y_i(x,z)=Y_i(x,z')$ for all $(x,z,z')$ (so one may denote the potential outcome as $Y_i(x)$), that the instrument is randomly assigned $Z_i\indep Y_i(\cdot),X_i(\cdot),$ and that treatment effects are constant, $Y_i(1)-Y_i(0)=\theta$ for all $i$. Under these assumptions $Y_i$ follows the linear model
$ Y_{i}=X_{i}\theta+\varepsilon_{i},$ where $\varepsilon_i=Y_i(0)$ and $E\left[Z_{i}\varepsilon_{i}\right]=0$.  Consequently, $\theta$ solves the moment condition \eqref{eq: moment conditions} for 
\[ Y\left(P\right)=E_{P}\left[Z_{i}Y_{i}\right],\,\,X\left(P\right)=E_{P}\left[Z_{i}X_{i}\right],
\]
which are (up to pre-multiplication by $E_P[Z_iZ_i']^{-1}$) equal to the reduced-form and first-stage coefficient vectors in the linear IV model, respectively.  $\triangle$

\paragraph{Example: Logit Model} As a second example, suppose that $D_i = (Y_i, X_i)$ for $Y_i \in \{0, 1\}$ a binary outcome and $X_i = (1, \tilde{X}_{i})\in\mathbb{R}^2$ an exogenous variable,  where $\tilde{X}_{i} \in \{x_1, ..., x_J\}$.  
If the researcher assumes a logistic regression (i.e. logit) model for $Y_i,$
\[ Y_i = 1\{X_i' \psi > \varepsilon_i\} \]
where $\varepsilon_i \sim \textrm{Logistic}(0,1)$ is independent of $X_i$, then under this model
\[ E_P[Y_i|X_i=x] = \Psi(x'\psi) \]
for $\Psi(u)=\frac{e^u}{1+e^u}$ the logistic function or, equivalently,
\[ \Psi^{-1}(E_P[Y_i|X_i=x]) = x'\psi \]
for $\Psi^{-1}(u)=\log\left(\frac{u}{1-u}\right)$ the logit function.  

We suppose that the object of interest $\theta\in\mathbb{R}^2$ parameterizes the conditional mean of  $Y$ given two as-yet-unobserved values of ${X}_{i},$
\[
\theta=(\theta_1,\theta_2)'=(\Psi^{-1}(E[Y_i|X_i=(1,x_1^*)]),\Psi^{-1}(E[Y_i|X_i=(1,x_2^*)]))',
\]
where  $x_1^*,x_2^*\not\in \{x_1, ..., x_J\}.$
The model implies that $\theta$ solves \eqref{eq: moment conditions} for 
\[ Y\left(P\right)=
        \begin{pmatrix}
        \Psi^{-1}(E_{P}\left[Y_i|X_i=(1,x_1)\right]) \\
        \vdots \\
        \Psi^{-1}(E_{P}\left[Y_i|X_i=(1,x_J)\right]) \end{pmatrix},
        \]
        \[
        X\left(P\right) = \begin{pmatrix}
            1 & x_1 \\
            \vdots & \vdots \\
            1 & x_J
        \end{pmatrix}
        \begin{pmatrix}
            \frac{x_2^*}{x_2^* - x_1^*} & -\frac{x_1^*}{x_2^* - x_1^*} \\
            -\frac{1}{x_2^* - x_1^*} & \frac{1}{x_2^* - x_1^*}
        \end{pmatrix}. ~~~ \triangle
        \] 

\subsection{Misspecification and Pseudo-True Values}\label{misspecification and pseudo-true values}

In many contexts researchers are concerned that their models may be misspecified.  
In the minimum distance setting we consider, this means that at the true $(P,\theta)$ pair  
\begin{equation}\label{eq: eta def}
g\left(\theta;P\right)=\eta\neq0,
\end{equation}
for $\eta$ a parameter that describes the impact of misspecification on the moments.
As the following examples highlight, we may have $\eta\neq 0$ for a variety of reasons.

\paragraph{Example: Linear IV (Continued)} Suppose we maintain the exclusion and independence assumptions for the instruments $Z_i,$ but allow treatment effects to be heterogeneous across units, $\text{Var}(Y_i(1)-Y_i(0))>0.$  If this treatment effect heterogeneity is correlated with heterogeneity in the first-stage effect $X_i(z)-X_i(z'),$ the results of \citet{imbens_identification_1994} imply that the linear IV moments are not in general zero when evaluated at the ATE $\theta.$  Instead, for $\beta$ the vector of one-instrument-at-a-time IV estimands (i.e. the IV coefficient using the first instrument by itself, the second by itself, and so on), $\iota\in\mathbb{R}^k$ the vector of ones, and $\circ$ the elementwise product, the implied value of $\eta$ is
\[ 
\eta = E_{P}\left[Z_{i}Y_{i}\right] - E_{P}\left[Z_{i}X_{i}\right] \theta = (\beta - \theta \cdot \iota) \circ E_P[Z_iX_i] \neq 0.
\]
Hence, the model is misspecified in the sense we consider whenever the one-at-a-time IV estimands differ from the ATE.  Note that the IV model can thus be misspecified even when we have only a single instrument, $k=1$: if in this case we further impose the \cite{imbens_identification_1994}
monotonicity assumption, the IV model will be misspecified if and only if the local average treatment effect (LATE) differs from the ATE. This is an instance of what \cite{andrews_purpose_2025} term econometric, rather than statistical, misspecification.

In this example we focus on misspecification arising from treatment effect heterogeneity, but our framework is sufficiently general to accommodate many other ways in which the researcher's assumptions could fail.  For instance, the exclusion restriction may fail, with $Y_i(x,z)\neq Y_i(x,z')$ for some $(x,z,z'),$ or independence may fail with $Z_i\not\indep (Y_i(\cdot),X_i(\cdot)).$ Each of these failures will imply a particular form for $\eta.$ $\triangle$

\paragraph{Example: Logit Model (Continued)} The logit model may be misspecified for a variety of reasons, for instance because the linear threshold model is incorrect and $X_i$ in fact enters nonlinearly, $Y_i=1\{h(X_i)>\varepsilon_i\},$ or because the linear threshold model is correct but the residual $\varepsilon_i$ does not follow a logistic distribution.  Whatever the reason for misspecification, we will have 
\[ \eta = \begin{pmatrix}
\Psi^{-1}\left(E_{P}\left[Y_i|X_i=(1,x_1)\right]\right) \\
\vdots \\
\Psi^{-1}\left(E_{P}\left[Y_i|X_i=(1,x_J)\right]\right) \end{pmatrix} - \begin{pmatrix}
    1 & x_1 \\
    \vdots & \vdots \\
    1 & x_J
\end{pmatrix}
        \begin{pmatrix}
            \frac{x_2^*}{x_2^* - x_1^*} & -\frac{x_1^*}{x_2^* - x_1^*} \\
            -\frac{1}{x_2^* - x_1^*} & \frac{1}{x_2^* - x_1^*}
        \end{pmatrix} \theta \neq 0, 
\] \\
where the conditional expectations $E_P[Y_i|X_i=(1,x_j)]$ will depend of the precise form of misspecification. $\triangle$

As these examples highlight, we consider settings where the true parameter $\theta$ is well-defined (e.g. based on some counterfactual or causal effect) even when the researcher's model is wrong in the sense that the true $\theta,P$ pair is incompatible with the model.
Of course, well-defined does not mean estimable, and if we allow $\eta\neq0$ and impose no other restrictions, identification of $\theta$ is hopeless since any
value of $\theta$ is compatible with any distribution $P$.  To avoid such a pessimistic conclusion, one route pursued in the literature, including in \citet{conley_plausibly_2012},
\citet{manski_how_2018}, \citet{masten_inference_2020}, \citet{armstrong_sensitivity_2021}, and \citet{rambachan_more_2023}, is to consider bounded relaxations of the model.  

To outline one approach along these lines, note that for a $P$-dependent positive-definite weighting matrix $W\left(P\right),$
the model is correctly specified if and only if the $W$-weighted norm of the misspecification parameter $\eta$ is equal to zero
\[
\left\Vert \eta\right\Vert_{W}=\sqrt{\eta'W\left(P\right)\eta} = 0.
\]
To allow the possibility of misspecification, the researcher could instead assume only that 
$\left\Vert \eta\right\Vert_{W}$ is bounded above by some known constant $d$.  The identified set for $\theta$ is then the set of values $\theta$ such that the minimum distance objective function
\[
Q_W(\theta;P)=\Vert g(\theta;P)\Vert_W^2
\]
takes a value no larger than $d^2$
\begin{equation}\label{eq: norm bound sets}
\Theta_{I}\left(P,d\right)=\left\{ \theta:Q_W(\theta;P)\le d^2\right\}.
\end{equation}

While this approach requires the researcher to specify the norm bound $d,$ the data do contain some information about this quantity.  Specifically, when $d^2<J_W(P)$ for 
$
J_W(P) = \min_{\theta}Q_W(\theta;P) 
$
the population analog of the $J$-statistic \citep{hansen_large_1982}, $\Theta_{I}\left(P,d\right)$ is empty, so the data reject the assumption that $\|\eta\|_W<d.$  Thus, as discussed in e.g. \citet{armstrong_sensitivity_2021} the data imply lower, but not upper, bounds on the degree of misspecification.

Another common practice when we are concerned with model misspecification is to focus on pseudo-true parameter values (c.f. \citealt{white_maximum_1982}, \citealt{muller_risk_2013},
\citealt{hansen_inference_2021}, and \citealt{andrews_misspecified_2023}),
 which are defined as the minimizers of a population objective function.
In our setting, the pseudo-true parameter corresponds to the value of $\theta$ at which $J_W(P)$ is attained, and is equal to the coefficient from a generalized least squares regression of $Y(P)$ on $X(P)$ weighting by $W(P),$
\[
\theta_{W}\left(P\right)=\argmin_{\theta}Q_W(\theta;P)=(X(P)'W(P)X(P))^{-1}X(P)'W(P)Y(P).
\] 

\paragraph{Example: Linear IV (Continued)}
In the linear IV model with treatment effect heterogeneity it is common to focus attention on the two-stage least squares (TSLS) estimand, which corresponds to the pseudo-true value using the TSLS weighting matrix $W(P) = E_P[Z_i Z_i']^{-1},$ and can be interpreted as a LATE under appropriate assumptions \citep{AngristImbens1995}. $\triangle$

\paragraph{Example: Logit Model (Continued)} In the logit model with misspecification, we cannot choose $\theta$ to match the full set of observed conditional means $E_P[Y_i|X_i=x].$  The weighting matrix governs how we prioritize matching different elements of this vector, and one option is to take $W(P)$ to be the diagonal matrix with $j$th diagonal element equal to the probability that $\tilde{X}_{i}=x_j$, $E_P[1\{\tilde{X}_{i}=x_j\}]$, which prioritizes matching the conditional mean for $X_i$ values which are more common in the population. $\triangle$
\\

The pseudo-true parameter corresponds to the identified set with $d^2=J_W(P),$ 
\[
\Theta_I(P,J_W(P)^{1/2})=\{\theta_{W}\left(P\right)\}.
\]
Hence, if a researcher assumes the true parameter value is equal to the pseudo-true parameter, this is the same as assuming that the degree of misspecification, measured in the norm $\|\cdot\|_W$, is as small as it can possibly be given the observed distribution $P$.\footnote{In just-identified case where $k=p,$ this corresponds to assuming $\eta=0$, so the model is correctly specified.}
If they instead allow the possibility that $\theta$ and $\theta_W(P)$ are different, then as discussed in the introduction it is not obvious how, if at all, the pseudo-true value $\theta_W(P)$ relates to the economic questions that motivate the analysis in the first place.  Consequently, it is unclear when we would want to estimate pseudo-true values.  The following two sections address this question by providing conditions under which optimal decisions depend on the data through (i) the population minimum distance objective $Q(\theta;P)$ and (ii) the pseudo-true value $\theta_W(P)$ in particular.

\section{Optimal Decisions and Minimum Distance}\label{sec: decision rules}

Researchers are often interested in estimating economic parameters in order to inform decisions by policymakers, businesses, or households. It is not obvious that pseudo-true values are suitable for this purpose.  To explore this question, we adopt a decision-theoretic perspective and ask under what conditions Bayesian decision-makers would be willing to base decisions on the minimum distance objective and pseudo-true value.  

\subsection{Decision Problem}

Consider a decision-maker who has to choose an action $a$ from a set of possible actions $\mathcal{A}$.  After choosing action $a,$ the decision-maker suffers a loss $L(a,\theta)$ that depends on the action taken and the true value for $\theta.$  If $\theta$ were known the optimal action for the decision-maker would be to simply choose $a\in\argmin_{a\in\mathcal{A}}L(a,\theta).$  

In fact $\theta$ is unknown and the decision-maker instead observes only the distribution $P\in\mathcal{P}$ of the data.  Hence, the decision-maker selects a decision rule $\delta:\mathcal{P}\to\mathcal{A}$ mapping data distributions into actions.  The decision-maker prefers rules $\delta$ that yield a lower loss, $L(\delta(P),\theta),$ but when $\theta$ cannot be uniquely determined based on $P$ (e.g. when $\theta$ is set-identified due to model misspecification), different decision rules $\delta$ will perform best at different $(P,\theta)$ pairs, and there generally will not be a uniformly best choice.

One way to choose $\delta$ is to consider Bayes decision rules, which weight losses
across different $(P,\theta)$ pairs according to a prior $\pi\in\Delta(\mathcal{P}\times\mathbb{R}^p)$. A Bayes decision rule $\delta_\pi$ minimizes the average loss under the prior,
\[
\delta_\pi \in \argmin_{\delta} \int L(\delta(P), \theta)d\pi(P, \theta).
\]
To compute $\delta_\pi$, it suffices to minimize the posterior expected loss at each $P,$
\[
\delta_\pi(P)\in \argmin_{a\in \mathcal{A}}\int L(a, \theta)\pi(\theta | P)d\theta,
\]
where we assume for simplicity that the posterior for $\theta|P$ is continuous and write $\pi(\theta | P)$ for the posterior density.

\paragraph{Example: Linear IV (Continued)}
As in \citet{andrews_model_2021}, suppose
the decision-maker needs to set a tax or subsidy $a\in\mathbb{R}$ for the treatment, where $a>0$ denotes a subsidy, while $a<0$ denotes a tax, and that the loss is $L(a,\theta)=(a-\theta)^2$.   That is, the optimal subsidy level is equal to the average treatment effect, while the loss increases quadratically as the subsidy departs from the ATE. This loss arises naturally when demand for treatment varies linearly with the subsidy amount.  $\triangle$

\subsection{Minimum-Distance Priors}

The posterior density $\pi(\theta|P)$ summarizes all decision-relevant information about the parameter $\theta$ given the data.  To 
connect minimum distance methods and Bayes decision rules, we consider a class of decision-makers for whom the minimum distance objective function is a sufficient statistic, in the specific sense that $\pi(\theta|P)$ is proportional to a function of $(Q_W(\theta;P),W(P),\theta).$  For such priors the minimum distance objective contains all $\theta$-relevant information and so is the natural basis for decision-making.
\begin{assumption}\label{Assumption: MD posterior}
The conditional prior $\pi(Y(P),\theta|X(P),W(P))$ is absolutely continuous with full support and a continuous density for all $X(P),W(P)$.  Moreover, for all $P\in\mathcal{P}$ and $\theta\in\Theta$
\[
\pi(\theta | P)= h(Q_W(\theta;P),W(P),\theta)c(P)
\]
for a non-negative function $h$ that is twice continuously differentiable in $Q$ and continuously differentiable in $\theta$, with a constant of proportionality $c(P)$ determined by $\int \pi(\theta|P)d\theta=1$.
\end{assumption}
The first part of the assumption is imposed for convenience and could be weakened.  The second part of the assumption connects the posterior distribution to the minimum distance objective, and is weaker than assuming $\pi(\theta|P)$ is proportional to a function $h(Q_W(\theta;P),\theta)$ as in the Gibbs posterior distributions studied in the statistics and machine learning literature (e.g. \citealt{catoni_pac-bayesian_2007}, \citealt{alquier_properties_2016}, \citealt{bissiri_general_2016}, \citealt{martin_chapter_2022}) and the quasi-Bayesian approach of \citet{chernozhukov_mcmc_2003}.
Under Assumption \ref{Assumption: MD posterior}, providing the decision-maker with $(Q_W(\cdot|P), W(P))$ is as good as providing them with the full data.  By contrast, when Assumption \ref{Assumption: MD posterior} fails the minimum distance objective fails to summarize the implications of $P$ for $\theta$ and minimum distance methods may be inappropriate.  Hence, we view Assumption \ref{Assumption: MD posterior} as a reasonable restriction in settings where Bayesian decision-markers are considering minimum distance methods.

Assumption \ref{Assumption: MD posterior} immediately implies that the posterior density $\pi(\theta|P)$ depends on the data only through $(W(P), X(P), Y(P))$.
\begin{lemma}\label{Lemma: Moment posterior}
Under Assumption \ref{Assumption: MD posterior}, $\pi(\theta | P)=\pi(\theta | W(P), X(P), Y(P))$ for all $P\in\mathcal{P}$.
\end{lemma}

\paragraph{Example: Linear IV (Continued)}
Focusing on the case where $W(P)$ is the two-stage least squares weighting matrix $W(P)=E_P[Z_iZ_i']^{-1},$ Lemma \ref{Lemma: Moment posterior} implies that the decision-maker's posterior for the ATE depends on the data only through the reduced-form and first stage regression coefficients, together with the covariance matrix of the instruments.  This rules out, for instance, situations where the decision-maker's beliefs about the ATE are informed by higher moments of $P$. $\triangle$ 
\\

 Assumption \ref{Assumption: MD posterior} further implies that the posterior density $\pi(\theta | P)$ can be expressed in terms of the minimum distance moments (\ref{eq: moment conditions}). In particular, the first part of Assumption \ref{Assumption: MD posterior} implies that the conditional priors for $\theta$ given $(W(P),X(P))$ and $\eta$ given $( W(P), X(P),\theta)$ are continuous. For brevity, let $\pi_\theta(\theta)=\pi(\theta|W(P),X(P))$ and $\pi_\eta(\eta| \theta)=\pi(\eta|W(P),X(P),\theta)$ denote their densities with respect to Lebesgue measure.
The posterior density of $\theta$ given $P$ is then
\begin{equation}\label{eq: moment representation of posterior}
\pi(\theta | P) = 
\frac{\pi_\theta(\theta)\pi_\eta(Y(P)-X(P)\theta | \theta)}{\int \pi_\theta(\theta)\pi_\eta(Y(P)-X(P)\theta | \theta)d\theta}=\frac{\pi_\theta(\theta)\pi_\eta(g(\theta;P)| \theta)}{\int \pi_\theta(\theta)\pi_\eta(g(\theta;P) | \theta)d\theta}.
\end{equation}
This resembles the posterior in a finite-sample problem with parameter $\theta,$ prior $\pi_\theta,$ and likelihood $\pi_\eta.$  Consistent with this resemblance, the posterior $\pi(\theta|P)$ will be non-degenerate with non-trivial uncertainty about the true value of $\theta$ even though the data distribution $P$ in our problem is perfectly known.  This reflects the fact that $\theta$ is not point-identified once $\eta$ is unknown, so when the conditional prior $\pi_\eta$ is non-dogmatic the decision-maker remains uncertain about $\theta$ even after observing $P.$

Under restrictions on the misspecification prior $\pi_\eta(\eta|\theta)$, Assumption \ref{Assumption: MD posterior} further implies that the prior density at $\eta$ conditional on $(W(P),\theta)$ depends only on $\eta'W(P)\eta$, $\pi_\eta(\eta|\theta)\propto f(\eta'W(P)\eta|\theta, W(P))$.
\begin{assumption}\label{Assumption: mean zero}
The conditional prior  $\pi_\eta(\eta|\theta)= \pi(\eta|\theta, W(P), X(P))$
\begin{itemize}
    \item[(i)] is strictly positive and twice continuously differentiable in $\eta$ and $\theta$;
    \item[(ii)] does not depend on $X(P)$: $\pi(\eta|\theta, W(P), X(P)) = \pi(\eta|\theta, W(P))$ for all $X(P)$; and
    \item[(iii)] has conditional mean zero given $W(P)$, $E_\pi[\eta | \theta,W(P)] = \int \eta \pi_\eta (\eta |\theta)d\eta = 0$.
\end{itemize}
Moreover, $\mathcal{P}$ is sufficiently rich that we can freely vary $(Y(P),X(P))$ while holding $W(P)$ fixed.
\end{assumption}
Assumption~\ref{Assumption: mean zero}(i) requires that the conditional prior on misspecification be sufficiently smooth.  Assumption~\ref{Assumption: mean zero}(ii) further requires that it not depend on the Jacobian matrix $X(P)$, and thus that $X(P)$ and $\eta$ be conditionally independent given $(\theta,W(P))$ under the prior.  In the context of linear IV, for instance, this requires that the researcher not update about the violation of the model's overidentifying restrictions based on the observed first stage.  Assumption~\ref{Assumption: mean zero}(iii) further requires that $\eta$ have conditional mean zero under the prior given $W(P),$ so the researcher thinks the model is ``right on average'' even after observing the weighting matrix.  To simplify the proof, we further assume that we can freely vary $(Y(P),X(P)).$
\begin{lemma}\label{Lemma: RII Prior}
Under Assumptions \ref{Assumption: MD posterior} and \ref{Assumption: mean zero}, the conditional prior density
$\pi_\eta(\eta | \theta)$ is rotationally
invariant with respect to $W(P)$, taking the form
\[\pi_\eta(\eta | \theta) = f(\eta'W(P)\eta| \theta, W(P)).\]
\end{lemma}
Lemma \ref{Lemma: RII Prior} implies that the prior density $\pi_\eta(\eta | \theta)$ is invariant to rotation of $W(P)^{\frac{1}{2}}\eta,$ in the sense that for any $\eta,$ $\tilde\eta$ such that $W(P)^{\frac{1}{2}}\eta=O W(P)^{\frac{1}{2}}\tilde\eta$ for an orthonormal matrix $O,$ the prior density is the same at $\eta$ and $\tilde\eta$, $\pi_\eta(\eta | \theta)=\pi_\eta(\tilde\eta | \theta)$.  The density $\pi_\eta(\eta | \theta)$ is thus constant on the ellipsoids $\{\eta:\eta'W(P)\eta=C\}$ for all $C$, from which it follows that $\pi_\eta(\eta | \theta)$
is an elliptically-contoured distribution (see \cite{muirhead_aspects_1982}, Chapter 1.5).

\paragraph{Example: Linear IV (Continued)}
Recall that $\eta=(\beta - \theta \cdot \iota) \circ E_P[Z_iX_i]$ measures the difference between the average treatment effect and the vector of one-instrument-at-a-time IV estimands.  One example of an elliptically-contoured distribution in this setting takes $\eta|\theta,W(P)\sim N(0,W(P)^{-1})$, which corresponds to $f(u|\theta, W(P))=\exp(-\frac{1}{2}u)$. Under this prior, each single-instrument estimand is equal to the ATE plus a mean-zero noise term, where the covariance matrix of the noise is determined by $W(P)$ and the first stage $E_P[Z_iX_i].$
There are many other rotation-invariant priors, however, including
a multivariate $t$ prior with $\nu$ degrees of freedom, where
$
f(u|\theta, W(P))= \left(1+\frac{1}{\nu}u\right)^{-\frac{\nu+k}{2}},
$
and an elliptically contoured power law which takes
$
f(u|\theta, W(P))= u^{-\kappa-1}
$
for $\kappa>k-1.$ $\triangle$ 
\\

Lemma \ref{Lemma: RII Prior} implies that under the conditions of Assumption  \ref{Assumption: mean zero} the posterior density $\pi(\theta|P)$ is proportional to a function of $(Q_W(\theta;P),W(P),\theta),$ as required by Assumption \ref{Assumption: MD posterior}, only if the (conditional) prior on $\eta$ is rotation-invariant.  In fact, rotation-invariance is both necessary and sufficient for Assumption \ref{Assumption: MD posterior} in settings where Assumption \ref{Assumption: mean zero} holds.

\begin{proposition}\label{Proposition: MD posterior}
Under Assumption \ref{Assumption: mean zero},  Assumption \ref{Assumption: MD posterior} holds if and only if for all $P\in\mathcal{P},$ the conditional prior $\pi(Y(P),\theta|X(P),W(P))$ is absolutely continuous and
\[
\pi(\theta | P) =
\frac{\pi_\theta(\theta)f(Q_W(\theta;P)|\theta, W(P))}{\int \pi_\theta(\theta)f(Q_W(\theta;P)|\theta, W(P))d\theta},
\]
for a non-negative function $f(u|\theta, W(P))$ with $\int f(\eta'\eta|\theta, W(P))d\eta < \infty.$
\end{proposition}
We have thus shown that, under the conditions on $\eta|\theta,W(P),X(P)$ imposed by Assumption \ref{Assumption: mean zero}, the minimum distance objective is sufficient in the sense of Assumption \ref{Assumption: MD posterior} if and only if our misspecification priors  $\pi_\eta$ are rotation invariant.  Hence, for such priors it is natural to focus on the minimum distance objective and, conversely, a focus on the minimum distance objective is most reasonable under such priors.  Justifying a focus on the pseudo-true value, i.e. on the argmin of the minimum distance objective, requires further restrictions, which we turn to next.

\section{Concentration-Inducing Priors}\label{sec: covergence-inducing priors}

As shown in the last section, for a fixed prior $\pi(\theta,P)$ satisfying Assumptions \ref{Assumption: MD posterior} and \ref{Assumption: mean zero} we obtain a non-degenerate posterior $\pi(\theta|P),$ and so are uncertain about the parameter $\theta$ even though the data distribution $P$ is perfectly known.  Plug-in decision rules based on the pseudo-true parameter value, by contrast, correspond to Bayes decision rules in the case where our posterior is a point-mass at $\theta_W(P).$
We next characterize sequences of priors satisfying Assumptions \ref{Assumption: MD posterior} and \ref{Assumption: mean zero} such that the corresponding posterior sequences concentrate at $\theta_W(P)$. Under such prior sequences the plug-in approach is correct in the limit, justifying a focus on the pseudo-true parameter.

While these prior sequences rationalize a focus on the pseudo-true value, we find them unsatisfactory in several respects.
First, these sequences assume that the model is misspecified with probability one (in the sense that $\|\eta\|_{W}>0$ almost surely under the prior), but take the expected \emph{magnitude} of misspecification to zero.  Hence, while these priors allow the possibility of misspecification, they assume that the degree of misspecification is arbitrarily small.
An assumption that the degree of misspecification is very small seems unreasonable in many economic applications.  Second, we show that concentration around the pseudo-true value is fragile in important respects.  If we take our concentration-inducing prior sequences and mix them, to an arbitrarily small degree, with any fixed full-support prior on $\eta|\theta$, concentration around the pseudo-true value immediately fails whenever $J_W(P)>0$.  Moreover, even under prior sequences which imply a vanishing degree of misspecification, concentration around the pseudo-true value requires that the prior on $\eta$ be sufficiently thin-tailed.

\subsection{Posterior Concentration}

Our sufficient conditions for posterior concentration require that the prior density for $W(P)^{-\frac{1}{2}}\eta$ be independent of $(\theta,W(P))$ and thin-tailed, in the sense that $f(\eta'W(P)\eta|\theta,W(P))$ decays at a faster-than-polynomial rate as $\eta'W(P)\eta\to\infty.$
\begin{assumption}\label{Assumption: Thin-Tailed Priors} For all $(W(P),\theta)$ and $f(\eta'W(P)\eta|\theta,W(P))$ as defined in Lemma \ref{Lemma: RII Prior}, $f(u|\theta,W(P))=f(u)$ where $f(u)$ is strictly positive, continuous and non-increasing in $u$, and satisfies $\lim_{u\to \infty}\frac{f(au)}{f(u)}=0$ for all $a>1$.
\end{assumption}
This assumption holds, for instance, when $\pi(\eta|\theta)$ is normal with variance proportional to $W(P)^{-1}$.
\paragraph{Example: Linear IV (Continued)}
If $\pi_{\eta}(\eta|\theta)$ is a $N(0,W(P)^{-1})$ density, then since $\exp\left(-\frac{a}{2}u\right)/\exp\left(-\frac{1}{2}u\right)=\exp\left(\frac{1-a}{2}u\right)$ and $a>1,$ Assumption \ref{Assumption: Thin-Tailed Priors} holds. $\triangle$ 
\\

Under sequences of priors that (i) satisfy Assumptions \ref{Assumption: MD posterior}, \ref{Assumption: mean zero} and \ref{Assumption: Thin-Tailed Priors} and (ii) take the degree of misspecification to be small a-priori, the posterior concentrates on the pseudo-true value. 
Formally, we consider a scale family of priors on the misspecification parameter $\eta,$ $\pi_{\eta,c}(\eta|\theta)\propto f(\frac{1}{c}\eta'W(P)\eta),$ where the scale parameter $c$ controls the magnitude of misspecification and the prior variance of $\eta$ scales with $c$.  Our main result in this section considers posterior behavior as the scale parameter becomes small, $c\to 0$, corresponding to priors that assume a vanishing degree of misspecification.

\begin{proposition}\label{Proposition: Posterior Concentration} Suppose
Assumptions \ref{Assumption: MD posterior}, \ref{Assumption: mean zero}, and \ref{Assumption: Thin-Tailed Priors} hold. For any continuous $\pi_\theta(\theta)$ with $\pi_\theta(\theta_W(P))>0,$ the posterior 
\[
\pi_c(\theta | P) = 
\frac{\pi_\theta(\theta) f(\frac{1}{c}Q_W(\theta;P))}{\int \pi_\theta(\theta) f(\frac{1}{c}Q_W(\theta;P)) d\theta}
\]
concentrates on $\theta_W(P)$ as $c\to 0$, in that for $B_\varepsilon(\theta_W(P))=\left\{\theta:\Vert \theta_W(P)-\theta \Vert <\varepsilon \right\}$,
\[
\lim_{c\to 0}\int 1\{\theta \notin B_\varepsilon(\theta_W(P))\}d\pi_c(\theta | P) = 0 \text{ for all }\varepsilon>0.
\] 
\end{proposition}

Proposition \ref{Proposition: Posterior Concentration} shows that for priors satisfying Assumptions \ref{Assumption: MD posterior}, \ref{Assumption: mean zero}, and \ref{Assumption: Thin-Tailed Priors} where the degree of misspecification is small, the  posterior distribution concentrates on $\theta_W(P).$  This is entirely expected when $J_W(P)=0$, since in this case the data provide no evidence of misspecification and priors with $c\to 0$ assume the model is nearly correct.  When $J_W(P)>0,$ by contrast, the data imply non-trivial misspecification but  Proposition \ref{Proposition: Posterior Concentration} shows that the posterior continues to concentrate.

Concentration of $\pi_c(\theta|P)$ implies convergence of the Bayes decision rule
\[
\delta_{\pi_c}(P)\in \argmin_{a\in \mathcal{A}}\int L(a, \theta)d\pi_c(\theta | P),
\]
under conditions on the decision problem:

\begin{proposition}\label{Proposition: Argmax theorem}
Suppose that $\mathcal{A}$ is compact under some metric $d$, that $\sup_{a,\theta}L(a,\theta)<\infty$, that $\sup_{\theta}|L(a,\theta)-L(a',\theta)|<\lambda\cdot d(a,a')$ for all $a, a'\in \mathcal{A}$ and some $\lambda>0$, that $L(a,\cdot)$ is continuous for all $a\in\mathcal{A}$, and that the loss $L(a,\theta)$ has a unique minimum for all $\theta$.  Then as $c\to 0,$ 
$
\delta_{\pi_c}(P)\to\argmin_{a\in\mathcal{A}}L(a,\theta_W(P)).
$ 
\end{proposition}

Proposition \ref{Proposition: Argmax theorem} shows that for bounded loss functions that are Lipschitz in $a$ and continuous in $\theta,$ Bayes decision rules corresponding to the priors we study converge to plug-in decision rules based on the pseudo-true value.  This result is useful for a number of reasons.  First, it shows that plug-in decision rules using the pseudo-true parameter value correspond to the limit of a sequence of Bayes decision rules for a large class of loss functions, providing one justification for such rules.  Second, it shows that the pseudo-true parameter value $\theta_W(P)$ is a sufficient statistic for communication with an audience whose priors take the limiting form we consider: a researcher looking to summarize the data for such an audience is justified in reporting only the pseudo-true parameter value, since it allows audience members to compute the optimal decision for whatever loss function they have, provided that loss satisfies the conditions of Proposition \ref{Proposition: Argmax theorem}.

The conditions in Proposition \ref{Proposition: Argmax theorem} are somewhat restrictive and rule out squared error loss on an unbounded domain. These conditions only are sufficient and not necessary, however, and we can obtain convergence of decision rules in many other settings by using additional structure for the loss function and prior.

\paragraph{Example: Linear IV (Continued)}
Proposition \ref{Proposition: Argmax theorem} does not apply in this example, because the loss $L(a,\theta)=(a-\theta)^2$ is neither bounded nor Lipschitz.  Nonetheless, if $\pi_\eta(\eta|\theta)$ corresponds to a $N\left(0,\left(\frac{1}{c}W(P)\right)^{-1}\right)$ distribution while  the prior on $\theta$ is flat, the posterior density is 
\begin{align*}
\pi_c(\theta | P) &= N\left(\theta_W(P), c\cdot(E_P[Z_iX_i]'W(P)E_P[Z_iX_i])^{-1}\right) \\  &= N\left(\theta_W(P), c\cdot(X(P)'W(P) X(P))^{-1}\right).
\end{align*}
Hence, for all $c$ the posterior distribution is a normal centered at $\theta_W(P)$ with variance proportional to $c$. Consistent with Proposition \ref{Proposition: Posterior Concentration}, this posterior converges weakly to a point mass at 
$\theta_W(P)$ as $c\to0$.  Moreover, despite the conditions of Proposition \ref{Proposition: Argmax theorem} not holding in this example, the Bayes decision rule $\delta_{\pi_c}(P)$ is equal to $\theta_W(P)$ for all $c.$ $\triangle$

\subsection{Posterior Concentration is Fragile}

While we have shown that the plug-in decision rule using the pseudo-true value corresponds to the limit of Bayes decision rules for certain sequences of priors, we think this justification carries limited weight, for multiple reasons.  For one the prior sequences used to establish convergence assume the degree of misspecification is negligible a-priori, which seems difficult to justify in many economic applications.  Moreover, as we next show the convergence established by Propositions \ref{Proposition: Posterior Concentration} and \ref{Proposition: Argmax theorem} is fragile in important respects.

\paragraph{Fragility to Prior Contamination}

If we mix the concentration-inducing priors studied in the previous section with any fixed, full support prior for $\eta|\theta$, posterior concentration fails when $J_W(P)>0.$
\begin{proposition}\label{Proposition: Posterior Non-Concentration}
Consider conditional priors of the form
\[
\pi_{\eta, c}^\phi(\eta | \theta) = (1-\phi)\pi_{\eta, c}(\eta | \theta)
+ \phi\pi^*_\eta(\eta | \theta)
\]
for any conditional prior $\pi^*_\eta(\eta|\theta)$ with strictly positive density and $\phi \in (0,1)$.
Suppose Assumption \ref{Assumption: Thin-Tailed Priors} holds.
If $J_W(P)>0$, then for any $\pi_\theta(\theta)$ the resulting posterior satisfies
\[
\lim_{c\to 0}\pi_c^\phi (\theta | P)
=
\frac{\pi_\theta(\theta)\pi_\eta^*(Y(P)-X(P)\theta| \theta)}{\int \pi_\theta(\theta)\pi_\eta^*(Y(P)-X(P)\theta| \theta)d\theta}.
\]
If instead $J_W(P)=0$, $\pi_\theta(\theta)$ is continuous, and $\pi_\theta(\theta_W(P))>0,$ then
\[
\lim_{c\to 0}\int 1\{\theta \notin B_\varepsilon(\theta_W(P))\}d\pi_c^{\phi}(\theta | P) = 0
\text{ for all }\varepsilon>0.
\]
\end{proposition}

Proposition \ref{Proposition: Posterior Non-Concentration} shows that if $J_W(P)>0$ and we contaminate the concentration-inducing prior $\pi_{\eta,c}(\eta|\theta),$ to an arbitrarily small extent, with any full-support prior $\pi_\eta^*$ for $\eta|\theta$ then the posterior converges to the same limit as if we had set $\pi_\eta(\eta|\theta)=\pi^*_\eta(\eta|\theta)$.  
By contrast, when $J_W(P)=0,$ the contaminated posterior continues to concentrate on $\theta_W(P)$ as $c\to 0.$  In this case the concentration-inducing component $\pi_{\eta,c}(\eta|\theta)$ receives asymptotically all of the posterior weight, despite the contamination by $\pi_\eta^*(\eta|\theta)$.

Proposition \ref{Proposition: Posterior Non-Concentration} can be interpreted in terms of pre-testing for model specification: in the case where $J_W(P)>0,$ the data imply that the model is non-trivially misspecified.  While the priors $\pi_{\eta,c}(\eta|\theta)$ imply that the model is misspecified with probability one, as $c\to 0$ they imply that the degree of misspecification is arbitrarily small. When we allow the possibility that $\eta$ is instead drawn from a fixed full-support distribution $\pi^*_\eta,$ for $c$ sufficiently small the data provide arbitrarily strong support for $\pi^*_\eta(\eta|\theta)$ over $\pi_{\eta,c}(\eta|\theta)$. 
Loosely speaking,  the concentration-inducing prior $\pi_{\eta,c}(\eta|\theta)$ is rejected in favor of the full-support prior $\pi^*_\eta(\eta|\theta)$.\footnote{We thank Jesse Shapiro for pointing out this connection.}

\paragraph{Fragility to Heavy Tails}

Even under prior sequences such that the degree of misspecification goes to zero, the concentration obtained in Proposition \ref{Proposition: Posterior Concentration} relies on thin tails for $\pi_\eta(\eta|\theta)$.  To illustrate this point, we show that posterior concentration around the pseudo-true value fails in two examples with heavy-tailed priors.

\paragraph{Example: Posterior Non-Concentration with $t$ Prior} 
Suppose that our prior on $\eta$ corresponds to a multivariate $t$ distribution centered at zero with scale matrix $W(P)^{-1}$ and $\tilde\nu$ degrees of freedom, $f(u) \propto \left(1 + \frac{1}{\tilde\nu}u\right)^{-\frac{\tilde\nu + k}{2}}$.
Provided $J_W(P)>0,$ if we define $\nu=\tilde\nu + k - p$ and 
\[
\Sigma(P) = J_W(P) \left({\nu} X(P)'W(P)X(P)\right)^{-1}
\]
one can show that 
\begin{align*}
    \lim_{c\to0} \pi_c(\theta | P) 
    &\propto \pi_\theta(\theta)\left(1 + {\nu}^{-1}(\theta - \theta_W(P))'\Sigma(P)^{-1}(\theta-\theta_W(P))\right)^{-({\nu}+p)/2},
\end{align*}
where the second term is the density for a multivariate $t$ distribution centered at $\theta_W(P)$, with scale matrix $\Sigma(P)$ and $\nu$ degrees of freedom.  Consequently, the $c\to0$ limiting posterior corresponds to updating the prior $\pi_\theta$ based on observing $\theta_W(P)\sim t_{{\nu}}(\theta,\Sigma(P)).$

Note that the degrees of freedom in the ``likelihood,'' $\nu$, is equal to the degrees of freedom in the misspecification prior $\pi_\eta$ plus the degree of over-identification, so a higher degree of over-identification leads to thinner tails for the posterior all else equal.  The scale parameter in the ``likelihood,'' $\Sigma(P) = J_W(P) \left({\nu} X(P)'W(P)X(P)\right)^{-1}$, is increasing in the $J$-statistic so cases where the moment conditions are observed to be more badly violated lead to a more uncertain posterior, all else equal. $\triangle$\\

The tail thickness of $f(\cdot)$ matters because it determines beliefs about the total level of misspecification conditional on a given value of the $J$-statistic.  To see this, let us again assume that $\eta|\theta\sim\pi_{\eta,c}(\eta|\theta)$ and consider the conditional distribution of $\|\eta\|_W$ conditional on the norm of $\eta$ exceeding a threshold $\tau$, $\|\eta\|_W|\left(\tau\leq\|\eta\|_W\right)$.  For thin-tailed priors (i.e. those satisfying Assumption \ref{Assumption: Thin-Tailed Priors}), one can show that for all $a>1$ and all positive constants $\tau>0$,
\begin{equation}\label{eq: conditional misspecification}
\lim_{c\to 0}Pr_{\pi_{\eta,c}}\left\{\|\eta\|_W\ge a\cdot \tau| \left(\tau\le \|\eta\|_W\right)\right\}\to0.
\end{equation}
Hence, even in the case where the model is known to be misspecified, as $c\to 0$ thin-tailed priors imply that the degree of \emph{additional} misspecification, beyond that implied by the lower bound $\tau$, is negligible.
Recall, however, that the $J$-statistic is itself a lower bound on the total degree of misspecification, $\|\eta\|_W\ge J_W(P)$.  Consistent with this, when $c\to 0$ thin-tailed priors imply that the total degree of misspecification must not be much larger than that suggested by the $J$-statistic, and thus that $\theta$ must be close to $\theta_W(P),$ the unique parameter value compatible with $\|\eta\|_W=J_W(P).$

By contrast, for $f$ corresponding to a multivariate $t$ distribution we have that
\begin{equation}\label{eq: conditional misspecification t}
\lim_{c\to 0}Pr_{\pi_{\eta,c}}\left\{\|\eta\|_W\ge a\cdot \tau| \left(\tau \le \|\eta\|_W \right)\right\}\to p(a).
\end{equation}
for a fixed, nonzero function $p(\cdot)$ that does not depend on $\tau$.  Consequently, in this case the researcher's belief about the total degree of misspecification is non-degenerate and, once $c$ is sufficiently small, scales proportionally with $\tau.$

While $t$-distributed priors for $\eta|\theta$ correspond to updates via a $t$ likelihood in the $c\to0$ limit, if we instead consider a multivariate power law prior for $\eta|\theta$ then dependence on $c$ vanishes entirely.

\paragraph{Example: Posterior Non-Concentration with Power Law Prior} 
Suppose $f(x) = x^{-\alpha}$ for $\alpha > k$.
Then for $\nu=2\alpha-p$, 
\[
\Sigma(P)=  J_W(P)\left(\nu X(P)'W(P)X(P)\right)^{-1},
\]
and all $c,$
\begin{align*}
    \pi_c(\theta | P) \propto \pi(\theta)\left(1 + \nu^{-1}(\theta-\theta_W(P))'\Sigma(P)^{-1}(\theta - \theta_W(P))\right)^{-(\nu + p)/2}
\end{align*}
which corresponds to the posterior distribution from observing $\theta_W(P)\sim t_\nu(\theta,\Sigma(P))$.
We again see that a higher degree of over-identification leads to thinner tails for the posterior, while a larger $J$-statistic leads to a more dispersed posterior.  $\triangle$\\

\section{Confidence Sets Based on Rotation-Invariance}\label{sec: confidence sets}

When $\pi(\theta|P)$ does not concentrate on the pseudo-true value, researchers will have non-trivial uncertainty about the true value of $\theta$ even after observing $P$, and the exact posterior will depend on the details of the prior.  Despite this dependence on the prior, we show that there nonetheless exist confidence intervals with correct ex-ante coverage for the true value $\theta$ under all priors satisfying Assumptions \ref{Assumption: MD posterior} and \ref{Assumption: mean zero}.  Since we previously argued that this class of priors has a close connection to minimum distance methods, we view these confidence sets as a natural summary for misspecification-driven uncertainty in settings where researchers adopt a minimum distance approach.

To state this result, suppose the researcher is interested in inference on a linear combination of the elements of $\theta,$ $v'\theta$ for $v\in\mathbb{R}^p.$  For $t_{k-p, 1-\frac{\beta}{2}}^*$ the level $1-\frac{\beta}{2}$ critical value for a standard
$t$ distribution with $k-p$ degrees of freedom, $\theta_{W}\left(P\right)=\argmin_{\theta}Q_W(\theta;P)$ the pseudo-true value, $J_W(P)=\min_\theta Q_W(\theta;P)$ the population $J$-statistic, and 
\[\sigma_v(P)=\sqrt{v'\left(X(P)'W(P)X(P)\right)^{-1}v}=
\sqrt{v'\left(\frac{1}{2}\frac{\partial^2}{\partial\theta\partial\theta'}Q_W(\theta;P)\right)^{-1}v}\]
a transformation of the Hessian of the population minimum distance objective, define the confidence interval 
\begin{equation}\label{eq: Rotation-invariance interval}
CI(P)=
\left[
v'\theta_W(P) \pm \sqrt{\frac{J_W(P)}{k-p}}\cdot \sigma_v(P)\cdot t_{k-p, 1-\frac{\beta}{2}}^*
\right].
\end{equation}
This interval has correct coverage conditional on $(W(P),X(P))$, and thus correct ex-ante coverage, under priors satisfying Assumptions \ref{Assumption: MD posterior} and \ref{Assumption: mean zero}.
\begin{proposition}\label{Proposition: Posterior Credible Sets with RII Prior}
For any prior $\pi$ such that Assumptions \ref{Assumption: MD posterior} and \ref{Assumption: mean zero} hold,
\[
Pr_\pi\left\{v'\theta\in CI(P)|W(P),X(P) \right\}=Pr_\pi\left\{v'\theta\in CI(P)\right\}=1-\beta.
\]
\end{proposition}

The confidence interval (\ref{eq: Rotation-invariance interval}) has a number of interesting features.  It is centered at the pseudo-true value and its width is governed by (i) the Hessian of the minimum distance objective function $\frac{\partial^2}{\partial\theta\partial\theta'}Q_W(\theta;P)$ and (ii) the population $J$-statistic $J_W(P).$  The fact that a smaller Hessian leads to wider confidence intervals resembles many other inference problems, though ours is unusual in that it is derived in a population setting and reflects uncertainty due to misspecification, rather than sampling uncertainty.  The dependence on the population $J$-statistic is also unusual, though intuitively reasonable given the connection to model misspecification.

The notion of coverage considered in Proposition \ref{Proposition: Posterior Credible Sets with RII Prior} is non-standard.  Since $P$ is observed in our setting, the frequentist coverage is either zero or one, $Pr_{(P,\theta)}\left\{v'\theta\in CI(P)\right\}\in\{0,1\}.$  Proposition \ref{Proposition: Posterior Credible Sets with RII Prior} instead considers average coverage under $\pi,$ 
\[
Pr_\pi\left\{v'\theta\in CI(P)\right\}=\int Pr_{(P,\theta)}\left\{v'\theta\in CI(P)\right\} d\pi(P,\theta),
\]
which measures the ex-ante coverage probability under the prior.  Proposition \ref{Proposition: Posterior Credible Sets with RII Prior} thus establishes that any Bayesian whose prior satisfies Assumption \ref{Assumption: MD posterior} thinks, before seeing the data, there is a $1-\beta$ probability that $CI(P)$ will cover $\theta$.\footnote{It is essential that Proposition \ref{Proposition: Posterior Credible Sets with RII Prior} uses the ex-ante probability.  Assumptions \ref{Assumption: MD posterior} and \ref{Assumption: mean zero} allow conflicting, near-dogmatic beliefs about $\theta$, so the only interval with non-zero worst-case ex-post coverage probability for $v'\theta$, $\inf_{\pi\in\Pi} Pr_\pi\left\{v'\theta\in CI(P)|P\right\}>0$ where $\Pi$ is the class of priors satisfying Assumptions \ref{Assumption: MD posterior} and \ref{Assumption: mean zero}, takes $CI(P)=\mathbb R$. Note further that this notion of coverage is often discussed in the literature on empirical Bayes (see, e.g., \cite{morris_parametric_1983,carlin_bayes_2000}.)}  In this sense, the interval $(\ref{eq: Rotation-invariance interval})$ has correct average coverage under any prior satisfying Assumption \ref{Assumption: MD posterior}.

We provide intuition for Proposition \ref{Proposition: Posterior Credible Sets with RII Prior} from two perspectives, first establishing a connection to generalized least squares and then showing that $CI(P)$ corresponds to a credible set under an improper power law prior.

\paragraph{Regression Interpretation}
Note that for 
\[
(\tilde{Y}, \tilde{X}, \tilde\eta) = W(P)^{\frac{1}{2}}(Y(P), X(P), \eta)
\]
we can write the minimum distance model allowing for misspecification, as 
\begin{equation}\label{eq: regression version}
\tilde{Y}(P) = \tilde{X}(P)\theta + \tilde{\eta},
\end{equation}
where under priors $\pi$ satisfying Assumptions \ref{Assumption: MD posterior} and \ref{Assumption: mean zero} the conditional distribution of $\tilde{\eta}|\tilde{X}(P),\theta$ is rotation-invariant, 
\[
\tilde{\eta}|\tilde{X}(P),\theta\sim O \tilde{\eta}|\tilde{X}(P),\theta
\] for all orthonormal matrices $O$.  Hence, for any prior satisfying Assumptions \ref{Assumption: MD posterior} and \ref{Assumption: mean zero} the problem of inference on $\theta$ conditional on $\tilde{X}(P)$ reduces to that of inference on the coefficient in a regression with a rotation-invariant error.  
However, the $t$-distribution for $t$-statistics holds whenever the error distribution is rotation-invariant, while (\ref{eq: Rotation-invariance interval}) is exactly the $t$-statistic confidence interval derived from (\ref{eq: regression version}) and so is valid under any prior satisfying Assumptions \ref{Assumption: MD posterior} and \ref{Assumption: mean zero}.\footnote{If we instead want a confidence interval for a multi-dimensional combination of the coefficients $\theta,$ the analogous approach based on $F$-statistics is also valid.}

\paragraph{Bayesian Interpretation}
Recall that  under the multivariate power-law prior for $\eta|\theta$, $f(x) = x^{-\alpha}$, the posterior $\pi_c(\theta|P)$ corresponds to a $t$ distribution with $2\alpha-p$ degrees of freedom.  Consequently, the confidence interval (\ref{eq: Rotation-invariance interval}) corresponds to a posterior credible set under a flat prior on $\theta$ and a multivariate power law prior on $\eta$ with $\alpha=k/2.$ This is an improper prior for $\eta,$ since in this case $\int \pi_\eta(\eta|\theta) d\eta=\infty,$ but yields a proper posterior  $\pi(\theta|P).$  Viewed from this perspective, Proposition \ref{Proposition: Posterior Credible Sets with RII Prior} shows that there exists an improper conditional prior on $\theta,\eta|W(P),X(P)$ whose credible sets have correct average coverage under all priors satisfying Assumptions \ref{Assumption: MD posterior} and \ref{Assumption: mean zero}.

\subsection{Comparison to Norm-Bound Identified Sets}

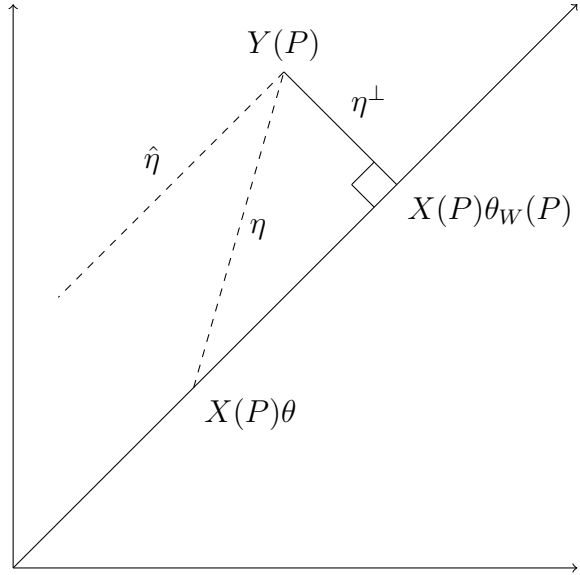
\begin{figure}[!h]
\centering
\begin{tikzpicture}[scale=3]
\draw[->] (0, 0) -- (2.5, 0);
\draw[->] (0, 0) -- (0, 2.5);

\draw[->] (0, 0) -- (2.5, 2.5);

\draw (1.7, 1.7) -- (1.2, 2.2);
\draw (1.45, 1.95) node[above right] {$\eta^\perp$};

\draw[dashed, -] (1.2, 2.2) -- (0.2, 1.2); 
\draw (0.7, 1.7) node[above left] {$\hat{\eta}$};

\draw (0.8, 0.8) node[below right] {$X(P)\theta$};

\draw[dashed](0.8, 0.8) -- (1.2, 2.2);
\draw (1.0, 1.5) node[right] {$\eta$};


\draw (1.6, 1.8) -- (1.5, 1.7);
\draw (1.5, 1.7) -- (1.6, 1.6);

\draw (1.7, 1.7) node[below right] {$X(P)\theta_W(P)$};
\draw (1.2, 2.2) node[above] {$Y(P)$};
\end{tikzpicture}
\caption{\footnotesize Relationship between the pseudo-true value and true $\theta$ in an example with scalar $\theta.$
The solid diagonal line represents the column space of $X(P)$.
The vector $\eta^\perp$ perpendicular to the column space captures the detectable component of the misspecification vector, while vector $\hat\eta$ parallel to the column space captures the undetectable component of misspecification.}
\label{fig:eta_components}
\end{figure}

We next compare the behavior of the confidence set (\ref{eq: Rotation-invariance interval}) to the identified set (\ref{eq: norm bound sets}) constructed under the bound $\|\eta\|_W\le d.$  To facilitate this comparison, note that
\[
\tilde \eta= W(P)^\frac{1}{2}\eta= M(P) \tilde\eta+(I-M(P))\tilde\eta =\hat\eta+\eta^\perp
\]
for 
\[
M(P)=\tilde X(P)(\tilde X(P)'\tilde X(P))^{-1}\tilde X(P)'
\]
the projection matrix onto $\tilde X(P)=W(P)^{\frac{1}{2}}X(P)$.  Here $\hat\eta$ and $\eta^\perp$ are the projection of $\tilde\eta$ onto the column span of $\tilde X(P)$ and the residual from this projection, respectively.  By definition $\|\eta^\perp\|^2=J_W(P)$, so the $J$-statistic measures the length of the projection residual, while $\hat\eta=\tilde{X}(P)(\theta_W(P)-\theta)$ governs the difference between the true and pseudo-true parameter values.  Intuitively, $\eta^\perp$ is the detectable component of the misspecification vector $\eta$, which has no effect on the bias of the pseudo-true value but governs the $J$-statistic.  Analogously, $\hat\eta$ is the undetectable component, which governs the bias but has no effect on the $J$-statistic. The overall degree of misspecification  reflects the sum of these terms, $\|\eta\|_W^2=\|\eta^\perp\|^2+\|\hat\eta\|^2$.  Figure \ref{fig:eta_components} visualizes this decomposition in a case where $\theta$ is scalar.

\begin{figure}
    \centering
    \begin{tikzpicture}
        \begin{groupplot}[
            group style={
                group size=2 by 1,
                horizontal sep=2cm
            },
            axis lines=middle,
            xmin=-0.3, xmax=2.5,
            ymin=-0.3, ymax=2.5,
            xtick=\empty,
            ytick=\empty,
            enlargelimits=false,
            axis equal image,
            axis line style={draw=none}, 
            clip=false,
        ]
        
        \nextgroupplot

        \draw[->] (0,0) -- (0,250);
        \draw[->] (0,0) -- (250,0) node[below] {$\theta$};

        \draw[->] (0,0) -- (250, 250);

        \draw (80, 180) circle [radius=70] node[above right] {\footnotesize $Y^A$};
        \draw (101, 159) circle [radius=70] node[above right] {\footnotesize $Y^B$};
        \node at (80, 180) [circle,fill,inner sep=1.0pt]{};
        \node at (101, 159) [circle,fill,inner sep=1.0pt]{};

        \draw[dashed](90, 90) -- (90, 0) node[below] {$l^B$};
        \draw[dashed](170, 170) -- (170, 0) node[below] {$u^B$};
        \draw[dashed](130, 130) -- (130, 0) node[below] {$\{\theta^A\}$};

        \nextgroupplot

        \draw[->] (0,0) -- (0,250);
        \draw[->] (0,0) -- (250,0) node[below] {$\theta$};

        \draw[->] (0,0) -- (250, 250);

        \draw (60, 60) -- (80, 180) node[above right] {\footnotesize $Y^A$} -- (200, 200);
        \draw (90, 90) -- (101, 159) node[above right] {\footnotesize $Y^B$} -- (170, 170);
        \node at (80, 180) [circle,fill,inner sep=1.0pt]{};
        \node at (101, 159) [circle,fill,inner sep=1.0pt]{};

        \draw[dashed](60, 60) -- (60, 0) node[below] {$l^A$};
        \draw[dashed](200, 200) -- (200, 0) node[below] {$u^A$};
        \draw[dashed](90, 90) -- (90, 0) node[below] {$l^B$};
        \draw[dashed](170, 170) -- (170, 0) node[below] {$u^B$};

        \end{groupplot}
    \end{tikzpicture}
    \caption{\footnotesize  Intervals for $\theta$ using the norm-bounding approach and the
rotation invariant prior approach. When dataset $A$ or $B$ is observed, the identified set or confidence interval for $\theta$ is given by $[l^A, u^A]$ or $[l^B, u^B]$, respectively, where in the left panel $l^A=u^A=\theta^A$.}
\label{fig:theta_identification}
    \label{fig:multipanel}
\end{figure}
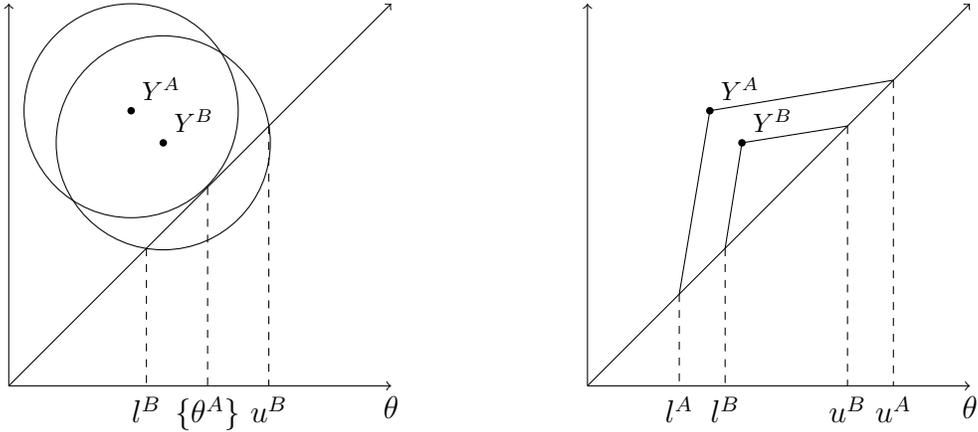

Equipped with this decomposition, note that for $\hat\eta(\theta,P)$
the undetectable misspecification component implied by $(\theta,P)$,
\[
\hat\eta(\theta,P)=W(P)^{\frac{1}{2}}X(P)(\theta_W(P)-\theta),
\]
we can write the identified set for $\theta$ as
\[\Theta_I(P,d)=\left\{\theta:\|\hat\eta(\theta,P)\|^2+J_W(P)\le d^2\right\}.
\]
This implies an identified set for $v'\theta$ equal to 
\[
\left[v'\theta_W(P)\pm \sigma_v(P)\sqrt{d^2-J_W(P)}\right].
\]
Hence, the bounds of the identified set correspond to values of $\theta$ which spend the misspecification ``budget'' $\|\hat\eta(\theta,P)\|^2\le d^2-J_W(P)$ obtained by subtracting the $J$-statistic from the a priori upper bound $d^2.$\footnote{One may recast this in Bayesian terms to parallel our earlier results by noting $\Theta_I(P,d)$ corresponds to the union of credible sets over priors $\pi$ such that $Pr_\pi\{\|\eta\|_W\le d\}=1$.}  As the degree of detectable misspecification becomes more severe, in the sense that the $J$-statistic grows larger, the length of the identified set shrinks.  The first panel of Figure \ref{fig:theta_identification} illustrates this, again focusing on the case where $\theta$ is scalar.  Here we hold $d$, $X(P),$ and $W(P)$ fixed but consider two possible values $Y(P),$ $Y^A$ and $Y^B,$ where $Y^A$ implies a larger $J$-statistic.  The identified set for $\theta$ is larger for $Y^B$ than $Y^A.$  Indeed, the $J$-statistic at $Y^A$ is exactly equal to $d,$ so the identified set collapses to the pseudo-true parameter value. Note that while the identified sets implied by the misspecification bounds in e.g. \cite{armstrong_sensitivity_2021} can exhibit these comparative statics, for inference that paper recommends ``fixed length'' confidence intervals which, as the name suggests, have fixed length and so avoid this issue.

In contrast to the identified set $\Theta_I(P,d)$, our proposed confidence intervals are wider for larger values of $J_W(P)$.  We can re-write the interval (\ref{eq: Rotation-invariance interval}) as 
\[
CI(P)=\left\{v'\theta:\|\hat\eta(\theta,P)\|\le t_{k-p,1-\frac{\beta}{2}}^*\frac{\|\eta^\perp(P)\|}{\sqrt{k-p}}\right\}.
\]
As the second panel of Figure \ref{fig:theta_identification} illustrates, the width of this interval is increasing in the size of the $J$-statistic, with a wider interval for $Y^A$ than for $Y^B.$  This seems like a potentially appealing property for uncertainty summaries in settings where researchers are concerned about misspecification.

\subsection{Confidence Intervals in Finite Samples}

So far, our results have abstracted from sampling uncertainty, considering the setting where $P$ is observed directly. While our analysis of estimands is fundamentally tied to this setting, our construction of confidence intervals extends naturally to settings with sampling uncertainty.

Formally, now suppose the researcher observes a finite sample $D^n=(D_1,...,D_n)$, and considers linear-in-parameters sample moment conditions
\[
g_n(\theta) = Y_n - X_n \theta 
\]
where $Y_n$, $X_n$, and $g_n(\theta)$ are the finite-sample counterparts of $Y(P)$, $X(P)$, and $g(\theta;P)$ respectively. At the true
$\theta$, define $\eta_n = g_n(\theta)$. 
In this setting, $\eta_n$
aggregates both population-level misspecification
and the sampling noise of the moments.
Assume the researcher's prior implies that $g_n(\theta)$ is continuously distributed, and thus nonzero almost surely.
Further define $W_n$ as the finite sample counterpart of $W(P)$ and $Q_{n}(\theta)= \Vert Y_n - X_n \theta \Vert_{W_n}^2$. We impose  analogs of Assumptions \ref{Assumption: MD posterior} and \ref{Assumption: mean zero}.

\begin{assumption}\label{Assumption: MD posterior finite}
The conditional prior $\pi(Y_n,\theta|X_n,W_n)$ is absolutely continuous with full support and a continuous density for all $X_n,W_n$.  Moreover, for all $D^n$ and $\theta\in\Theta$
\[
\pi(\theta | D^n)= h(Q_{n}(\theta),W_n,\theta)c(D^n)
\]
for a non-negative function $h$ that is twice continuously differentiable in $Q$ and continuously differentiable in $\theta$, with a constant of proportionality $c(D^n)$.
\end{assumption}

\begin{assumption}\label{Assumption: mean zero finite}
The conditional prior  $\pi_{\eta_n}(\eta_n|\theta)=\pi(\eta_n|\theta, W_n, X_n)$:
\begin{itemize}
    \item[(i)] is strictly positive and twice continuously differentiable in $\eta_n$ and $\theta$;
    \item[(ii)] does not depend on $X_n$: $\pi(\eta_n|\theta, W_n, X_n) = \pi(\eta_n|\theta, W_n)$ for all $X_n$; and
    \item[(iii)] has conditional mean zero given $W_n$, $E_\pi[\eta_n | \theta,W_n] = \int \eta_n \pi_{\eta_n} (\eta_n |\theta)d\eta_n = 0$.
\end{itemize}
Moreover, the support of the data is such that we can freely vary $(X_n,Y_n)$ while holding $W_n$ fixed.
\end{assumption}

Under these conditions we obtain a finite-sample analog of Proposition \ref{Proposition: Posterior Credible Sets with RII Prior}.  Define
\[
CI_n=
\left[
v'\theta_{n,W_n} \pm \sqrt{\frac{J_{n,{W_n}}}{k-p}}\cdot \sigma_{v,n}\cdot t_{k-p, 1-\frac{\beta}{2}}^*
\right],
\]
where $\theta_{n,W_n},$ $J_{n,{W_n}},$ and  $\sigma_{v,n}$ are the finite-sample counterparts of $\theta_{W}(P)$, $J_W(P)$, and $\sigma_{v}(P)$ respectively.

\begin{proposition}\label{Proposition: Posterior Credible Sets with RII Prior finite}
For any prior $\pi$ such that Assumptions \ref{Assumption: MD posterior finite} and \ref{Assumption: mean zero finite} hold,
\[
Pr_\pi\left\{v'\theta\in CI_n|W_n,X_n \right\}=Pr_\pi\left\{v'\theta\in CI_n\right\}=1-\beta.
\]
\end{proposition}

\subsection{Confidence Intervals under Local Misspecification}

The last section showed that our confidence sets for linear-in-parameters moment condition models with $P$ observed directly translate to finite-sample results, again for linear-in-parameters models.  Standard arguments imply, however, that nonlinear moment condition models are approximately linear in large samples, so long as the degree of misspecification is sufficiently small.  Thus, our confidence sets similarly extend to locally misspecified, but potentially nonlinear, models.

To sketch this extension, following \cite{armstrong_sensitivity_2021} let us suppose that the sample of size $n$ is generated from a DGP $P_n$ where the pair $(P_n,\theta)$ satisfy the local misspecification condition\footnote{See also, e.g., \cite{armstrong_adapting_2025} for related work on local misspecification.}
\[
g(\theta; P_n) = \mu/\sqrt{n}.
\]
\cite{armstrong_sensitivity_2021} show that under local misspecification and additional regularity conditions, the problem of 
 constructing a confidence interval for a scalar parameter $\kappa(\theta)$ (for some possibly nonlinear function $\theta \mapsto \kappa(\theta)$) is asymptotically equivalent to the problem of constructing a confidence interval for the parameter $K\theta $ (for $1\times p$ derivative matrix of $\kappa$ at true $\theta$ given by $K$) in the approximately linear model
\[
Y_L = -\Gamma_L \theta + \mu + \varepsilon, \quad \varepsilon \sim N(0,\Sigma)
\]
where $\Gamma_L$ corresponds to the Jacobian of the moments at the true parameter value and we observe $Y_L$ and $\Gamma_L,$ while $\Sigma$ is known.\footnote{Local misspecification plays an important role in this result, not only allowing linearization of nonlinear moment conditions but also ensuring that the asymptotic variance of GMM estimators remains of the usual form, unlike in the globally misspecified cases considered by e.g.  \citet{hall_large_2003} and \citet{hansen_inference_2021}.} 

Defining $\eta = \mu + \varepsilon$, if our prior on $\eta$ is rotation invariant (as follows, for instance, if we have a $N(0,\Omega)$ prior on $\mu$ with $\Omega+\Sigma\propto W^{-1}$), then the limit experiment is isomorphic
to a linear regression with rotation-invariant errors. Specifically, for $X_L = -\Gamma_L$, we can re-write the model as
\[
Y_L = X_L \theta + \eta.
\]
We now state the natural analogs of Assumptions \ref{Assumption: MD posterior} and \ref{Assumption: mean zero} for this setting, denoting the data in the limit experiment as $D_L= (Y_L,W_L, X_L)$.

\begin{assumption}\label{Assumption: MD posterior local}
The conditional prior $\pi(Y_L,\theta|X_L,W_L)$ is absolutely continuous with full support and a continuous density for all $X_L,W_L$.  Moreover, for all $D_L$ and $\theta\in\Theta$
\[
\pi(\theta | D_L)= h(Q_{W_L}(\theta;D_L),W_L,\theta)c(D_L)
\]
for a non-negative function $h$ that is twice continuously differentiable in $Q$ and continuously differentiable in $\theta$, with a constant of proportionality $c(D_L)$ determined by $\int \pi(\theta|D_L)d\theta=1$.
\end{assumption}

\begin{assumption}\label{Assumption: mean zero local}
The conditional prior  $\pi_\eta(\eta|\theta)=\pi(\eta|\theta, W_L, X_L)$:
\begin{itemize}
    \item[(i)] is strictly positive and twice continuously differentiable in $\eta$ and $\theta$;
    \item[(ii)] does not depend on $X_L$: $\pi(\eta|\theta, W_L, X_L) = \pi(\eta|\theta, W_L)$ for all $X_L$; and
    \item[(iii)] has conditional mean zero given $W_L$, $E_\pi[\eta | \theta,W_L] = \int \eta \pi_\eta (\eta |\theta)d\eta = 0$.
\end{itemize}
Moreover, we can freely vary $(X_L,Y_L)$ while holding $W_L$ fixed.
\end{assumption}

Constructing a confidence interval for scalar $K\theta$ is then accomplished by defining 
\[\sigma_{K,L}=\sqrt{K\left(X_L'W_L X_L\right)^{-1}K'}, \quad CI_L=
\left[
K\theta_{L,W_L} \pm \sqrt{\frac{J_{L,W_L}}{k-p}}\cdot \sigma_{K,L}\cdot t_{k-p, 1-\frac{\beta}{2}}^*
\right],\]
for $\theta_{L,W_L},J_{L,W_L},$ and $\sigma_{K,L}$ the limit experiment counterparts of $\theta_W(P),J_W(P)$, and $\sigma_v(P),$ respectively.

\begin{proposition}\label{Proposition: Posterior Credible Sets with RII Prior local}
For any prior $\pi$ such that Assumptions \ref{Assumption: MD posterior local} and \ref{Assumption: mean zero local} hold
\[
Pr_\pi\left\{K\theta\in CI_L|W_L,X_L \right\}=Pr_\pi\left\{K\theta\in CI_L\right\}=1-\beta.
\]
\end{proposition}

\section{Conclusion}

We study the problem of inference in misspecified linear minimum distance models, and show that under restrictions on the prior (i) a Bayesian decision-maker is content to summarize the data using the minimum distance objective if and only if their prior on model misspecification satisfies a rotation-invariance condition; (ii) a Bayesian is content to further summarize the data via the minimum distance estimand---the pseudo-true value---if their prior on misspecification is thin-tailed and has variance going to zero; and (iii) for any Bayesian with a rotation-invariant prior as in (i), even if the conditions of (ii) don't hold, a particular confidence set, centered at the pseudo-true value and with width proportional to the square root of a population $J$-statistic, ensures correct coverage for the true parameter value with no restriction on the magnitude of misspecification.

Informative inference is possible because rotation-invariant priors restrict the ``direction'' of misspecification, controlling the impact of misspecification on the pseudo-true value relative to the impact on the $J$-statistic. Our key assumption is that researchers' choice of weighting matrix reflects their beliefs about the relative likelihood of different kinds of misspecification, which appears consistent with the way in which some researchers already choose their weighting matrices.\footnote{For instance, \citet{benhabib_wealth_2019} write ``The weighting matrix $W$ in the baseline is a diagonal matrix with identical weights for all but the last moment of both the wealth distribution and the mobility moments, which are overweighted (ten times), according to the prior that matching the tail of the distribution is a fundamental objective of our exercise.''} For such researchers, our work suggests that a natural way to account for the possibility of model misspecification is by implementing confidence intervals of the form in Section \ref{sec: confidence sets}, with width that scales with the (population or finite-sample) $J$-statistic.

\clearpage

\begin{spacing}{1.15}
\bibliographystyle{ecta}
\bibliography{paper}
\end{spacing}

\section{Proofs}

\paragraph{Proof of Lemma \ref{Lemma: Moment posterior}}
$Q_W(\cdot | P)$
can be expressed as a function of $(X(P), W(P), Y(P))$,
so by Assumption \ref{Assumption: MD posterior},
\begin{align*}
    \pi(\theta \mid P)
    &=
    \pi(\theta \mid Q_W(\cdot ; P), W(P))\\
    &=
    \pi(\theta \mid X(P), W(P), Y(P))
\end{align*}
as we aimed to show. $\square$

\paragraph{Proof of Lemma \ref{Lemma: RII Prior}}
Throughout the proof, we fix an arbitrary positive-definite matrix $W$ and suppress it from the notation; all densities are implicitly conditioned on $W$.  By Assumption~\ref{Assumption: mean zero}(ii), $\pi_\eta(\eta|\theta)$ does not depend on $X$, and we write $\psi(\eta,\theta) = \ln \pi_\eta(\eta|\theta)$.

Fix any full-rank $X$ and let $Y = X\theta + \eta$.  By Bayes' theorem,
\[
\pi(Y|\theta,W,X) = \frac{\pi(\theta|Y,W,X)\cdot \pi(Y|W,X)}{\pi(\theta|W,X)}.
\]
Since $\pi(Y|\theta,W,X) = \pi_\eta(Y - X\theta|\theta)$ by a change of variables, and $\pi(\theta|Y,W,X) = h(Q_W(\theta;P),W,\theta)\cdot c(P)$ by Assumption~\ref{Assumption: MD posterior} and Lemma~\ref{Lemma: Moment posterior}, we obtain
\begin{equation}\label{eq: log decomposition}
\psi(\eta,\theta) = \underbrace{\ln h(\eta'W\eta,\,W,\,\theta)}_{A(\eta'W\eta,\,\theta)} \;+\; \underbrace{\ln\bigl[c(P)\cdot \pi(Y|W,X)\bigr]}_{B(Y,X)} \;+\; \underbrace{\bigl(-\ln \pi(\theta|W,X)\bigr)}_{S(\theta,X)},
\end{equation}
where $A(\eta'W\eta,\theta)$ depends on $\eta$ only through $\eta'W\eta$ and does not depend on $X$; $B(Y,X)$ depends on $Y$ and $X$ but not on $\theta$; and $S(\theta,X)$ depends on $\theta$ and $X$ but not on $Y$.  Since the left-hand side $\psi(\eta,\theta)$ does not depend on $X$ by Assumption~\ref{Assumption: mean zero}(ii) and $A$ does not depend on $X$ since $h$ takes arguments $(Q_W, W, \theta)$, it follows that $B(Y,X) + S(\theta,X)$ does not depend on $X$.

We now take mixed partial derivatives to eliminate the nuisance terms $B$ and $S$.  All composed functions are sufficiently smooth that the order of differentiation may be interchanged freely, by Assumptions~\ref{Assumption: MD posterior} and~\ref{Assumption: mean zero}(i).  Differentiating \eqref{eq: log decomposition} with respect to $\theta$ (holding $Y,X,W$ fixed, so $\eta = Y - X\theta$ varies with $\theta$) and then with respect to $Y'$ (holding $\theta,X,W$ fixed) eliminates $B$ and $S$: the term $\nabla_\theta B = 0$ since $B$ does not depend on $\theta$, and $\frac{\partial}{\partial Y'}\nabla_\theta S = 0$ since $S$ does not depend on $Y$.  The resulting identity is
\begin{equation}\label{eq: mixed partial identity}
-X'\nabla^2_{\eta\eta}\psi(\eta,\theta) + \nabla^2_{\eta\theta}\psi(\eta,\theta) = -X'G(\eta,\theta) + H(\eta,\theta),
\end{equation}
where
\[
G(\eta,\theta) = 4\frac{\partial^2 A}{\partial Q^2}\,W\eta\eta'W + 2\frac{\partial A}{\partial Q}\,W, \qquad
H(\eta,\theta) = \left(\nabla_\theta \frac{\partial A}{\partial Q}\right)2\eta'W.
\]

Since $\psi$ does not depend on $X$ by Assumption~\ref{Assumption: mean zero}(ii), $\nabla^2_{\eta\eta}\psi$ and $\nabla^2_{\eta\theta}\psi$ do not depend on $X$.  Likewise, $G$ and $H$ do not depend on $X$, since $A = \ln h(Q,W,\theta)$ and $Q = \eta'W\eta$ do not involve $X$.  Defining $M = \nabla^2_{\eta\eta}\psi - G$ and $N = \nabla^2_{\eta\theta}\psi - H$, neither of which depends on $X$, the identity becomes $-X'M + N = 0$ for all $X$ in the open set.  For any $X_0$ in this set and any sufficiently small perturbation $\Delta X$ with $X_0 + \Delta X$ still in the set, $(\Delta X)'M = 0$.  Taking $\Delta X = \varepsilon\, e_j e_i'$ for each $j\in\{1,\dots,k\}$ and $i\in\{1,\dots,p\}$ gives $(\Delta X)'M = \varepsilon\, e_i\, e_j'\, M = 0$; since this is the outer product of the nonzero vector $e_i$ with the row vector $e_j'M$, we obtain $e_j'M = 0$ for all $j$, so $M = 0$ and $N = X_0'M = 0$.

The equality $M = 0$ means $\nabla^2_{\eta\eta}\psi(\eta,\theta) = \nabla^2_{\eta\eta}A(\eta'W\eta,\theta)$, so $\nabla^2_{\eta\eta}[\psi(\eta,\theta) - A(\eta'W\eta,\theta)] = 0$ for all $\eta\in\mathbb{R}^k$.  A function whose Hessian is identically zero on $\mathbb{R}^k$ is affine, so there exist $\lambda_0(\theta)$ and $\lambda_1(\theta)\in\mathbb{R}^k$ such that
\[
\psi(\eta,\theta) = A(\eta'W\eta,\theta) + \lambda_0(\theta) + \lambda_1(\theta)'\eta.
\]
Exponentiating, $\pi_\eta(\eta|\theta) = g(\eta;\theta)\cdot \exp(\lambda_1(\theta)'\eta)$ where $g(\eta;\theta) = \exp(A(\eta'W\eta,\theta) + \lambda_0(\theta))$ depends on $\eta$ only through $\eta'W\eta$ and is therefore an even function, $g(-\eta;\theta) = g(\eta;\theta)$.

It remains to show $\lambda_1(\theta) = 0$.  By Assumption~\ref{Assumption: mean zero}(iii),
\[
0 = \int_{\mathbb{R}^k} \eta\, g(\eta;\theta)\, \exp(\lambda_1(\theta)'\eta)\, d\eta.
\]
Pre-multiplying by $\lambda_1(\theta)'$ gives
\[
0 = \int_{\mathbb{R}^k} (\lambda_1(\theta)'\eta)\, g(\eta;\theta)\, \exp(\lambda_1(\theta)'\eta)\, d\eta.
\]
Decomposing $\mathbb{R}^k$ into $S_+ = \{\eta : \lambda_1'\eta > 0\}$ and $S_- = \{\eta : \lambda_1'\eta < 0\}$ (the hyperplane $\{\lambda_1'\eta = 0\}$ has measure zero) and substituting $u = -\eta$ in the integral over $S_-$, using $g(-u;\theta) = g(u;\theta)$,
\[
0 = \int_{S_+} (\lambda_1'\eta)\, g(\eta;\theta)\, \bigl[e^{\lambda_1'\eta} - e^{-\lambda_1'\eta}\bigr]\, d\eta.
\]
For $\eta\in S_+$ and $\lambda_1\neq 0$: $\lambda_1'\eta > 0$, $g(\eta;\theta) > 0$ by Assumption~\ref{Assumption: mean zero}(i), and $e^{\lambda_1'\eta} - e^{-\lambda_1'\eta} > 0$ since the exponential function is strictly increasing.  The integrand is strictly positive on $S_+$, which has positive measure when $\lambda_1\neq 0$, contradicting the integral being zero.  Hence $\lambda_1(\theta) = 0$ and
\[
\pi_\eta(\eta | \theta) = f(\eta'W\eta| \theta, W) = \exp(A(\eta'W\eta, \theta) + \lambda_0(\theta)).
\]
Restoring the suppressed $W(P)$-dependence completes the proof. $\square$

\paragraph{Proof of Proposition \ref{Proposition: MD posterior}}

Immediate from Equation (\ref{eq: moment representation of posterior}) and Lemma \ref{Lemma: RII Prior}. $\Box$

\paragraph{Proof of Proposition \ref{Proposition: Posterior Concentration}}
Note that for the claim to hold, it is necessary and sufficient that for $\tilde{W}(P)=X(P)'W(P)X(P)$ and 
\[
\tilde{B}_\varepsilon(\theta_W(P))=\left\{\theta:\Vert \theta_W(P)-\theta \Vert_{\tilde{W}(P)} <\varepsilon \right\},
\]
we have that for all $\varepsilon>0,$
\begin{align*}
    \int 1\{\theta \in \tilde{B}_\varepsilon(\theta_W(P))\}d\pi_c(\theta\mid P)
    &=\frac{\int 1\{\theta \in \tilde{B}_\varepsilon(\theta_W(P))\}f\left(\frac{1}{c}Q_W(\theta ; P)\right)\pi_\theta(\theta)d\theta}{\int f\left(\frac{1}{c}Q_W(\theta ; P)\right) \pi_\theta(\theta)d\theta}\to 1.
\end{align*}
Taking the inverse of this probability (which is possible because the posterior assigns strictly positive mass to neighborhoods of the pseudo-true value) yields
\begin{equation}\label{eq: inverse probability}
    1+\frac{\int 1\{\theta \notin \tilde{B}_\varepsilon(\theta_W(P))\}f\left(\frac{1}{c}Q_W(\theta ; P)\right)\pi_\theta(\theta)d\theta}{\int 1\{\theta \in \tilde{B}_\varepsilon(\theta_W(P))\}f\left(\frac{1}{c}Q_W(\theta ; P)\right) \pi_\theta(\theta)d\theta},
\end{equation}
where to prove the result it suffices to show that the second term goes to zero.

To this end, note that for any $a>1$, we can re-write the second term of (\ref{eq: inverse probability}) as
\[
\frac{\int (1\{\theta \in \tilde{B}_{a\varepsilon}(\theta_W(P))\setminus \tilde{B}_{\varepsilon}(\theta_W(P))\} + 1\{\theta \notin \tilde{B}_{a\varepsilon}(\theta_W(P))\})f\left(\frac{1}{c}Q_W(\theta ; P)\right)\pi_\theta(\theta)d\theta}{\int 1\{\theta \in \tilde{B}_\varepsilon(\theta_W(P))\}f\left(\frac{1}{c}Q_W(\theta ; P)\right) \pi_\theta(\theta)d\theta}.
\]
Note, however, that we can express the minimum distance objective as
\[
Q_W(\theta ; P)
= J_W(P) + 
\|\theta - \theta_W(P)\|_{\tilde{W}(P)}^2.
\]
Thus, since we have assumed $f$ is non-increasing
\[
\frac{\int1\{\theta \notin \tilde{B}_{a\varepsilon}(\theta_W(P))\}f\left(\frac{1}{c}Q_W(\theta ; P)\right)\pi_\theta(\theta)d\theta}{\int 1\{\theta \in \tilde{B}_\varepsilon(\theta_W(P))\}f\left(\frac{1}{c}Q_W(\theta ; P)\right) \pi_\theta(\theta)d\theta}
\le
\]
\[
\frac{\int1\{\theta \notin \tilde{B}_{a\varepsilon}(\theta_W(P))\}f\left(\frac{1}{c}(J_W(P) + a^2\varepsilon^2))\right)\pi_\theta(\theta)d\theta}{\int 1\{\theta \in \tilde{B}_\varepsilon(\theta_W(P))\}f\left(\frac{1}{c}(J_W(P) + \varepsilon^2)\right) \pi_\theta(\theta)d\theta}=
\]
\[
\frac{f\left(\frac{1}{c}(J_W(P) + a^2\varepsilon^2)\right)}{f\left(\frac{1}{c}(J_W(P) + \varepsilon^2)\right)}
\frac{\int1\{\theta \notin \tilde{B}_{a\varepsilon}(\theta_W(P))\}\pi_\theta(\theta)d\theta}{\int 1\{\theta \in \tilde{B}_\varepsilon(\theta_W(P))\} \pi_\theta(\theta)d\theta},
\]
where the first term converges to zero as $c\to 0$ by Assumption \ref{Assumption: Thin-Tailed Priors}, while the second doesn't depend on $c$.  Hence, 
\[
\lim_{c\to 0}\frac{\int1\{\theta \notin \tilde{B}_{a\varepsilon}(\theta_W(P))\}f\left(\frac{1}{c}Q_W(\theta ; P)\right)\pi_\theta(\theta)d\theta}{\int 1\{\theta \in \tilde{B}_\varepsilon(\theta_W(P))\}f\left(\frac{1}{c}Q_W(\theta ; P)\right) \pi_\theta(\theta)d\theta}
=0.
\]
Note, next, that 
\[
\frac{\int 1\{\theta \in \tilde{B}_{a\varepsilon}(\theta_W(P))\setminus \tilde{B}_{\varepsilon}(\theta_W(P))\} f\left(\frac{1}{c}Q_W(\theta ; P)\right)\pi_\theta(\theta)d\theta}{\int 1\{\theta \in \tilde{B}_\varepsilon(\theta_W(P))\}f\left(\frac{1}{c}Q_W(\theta ; P)\right) \pi_\theta(\theta)d\theta}\le
\]
\[
\frac{\int 1\{\theta \in \tilde{B}_{a\varepsilon}(\theta_W(P))\setminus \tilde{B}_{\varepsilon}(\theta_W(P))\} f\left(\frac{1}{c}(J_W(P) + \varepsilon^2)\right)\pi_\theta(\theta)d\theta}{\int 1\{\theta \in \tilde{B}_\varepsilon(\theta_W(P))\}f\left(\frac{1}{c}(J_W(P) + \varepsilon^2)\right) \pi_\theta(\theta)d\theta}=\]
\[
\frac{\int 1\{\theta \in \tilde{B}_{a\varepsilon}(\theta_W(P))\setminus \tilde{B}_{\varepsilon}(\theta_W(P))\} \pi_\theta(\theta)d\theta}{\int 1\{\theta \in \tilde{B}_\varepsilon(\theta_W(P))\} \pi_\theta(\theta)d\theta},
\]
where the last expression goes to zero as we take $a\to1.$  Together with our earlier argument this implies that 
\begin{equation}
    \lim_{c\to 0}\frac{\int 1\{\theta \notin \tilde{B}_\varepsilon(\theta_W(P))\}f\left(\frac{1}{c}Q_W(\theta ; P)\right)\pi_\theta(\theta)d\theta}{\int 1\{\theta \in \tilde{B}_\varepsilon(\theta_W(P))\}f\left(\frac{1}{c}Q_W(\theta ; P)\right) \pi_\theta(\theta)d\theta}=0,
\end{equation}
and so completes the proof. $\Box$

\paragraph{Proof of Proposition \ref{Proposition: Argmax theorem}}

For each $c$, define
$m_c(a)=\int L(a,\theta)d\pi_c(\theta|P)$
and let
$m(a)=L(a,\theta_W(P)).$
Since Proposition \ref{Proposition: Posterior Concentration} implies that $\pi_c(\cdot|P)$ converges weakly to a point mass at $\theta_W(P)$, and since $L(a,\cdot)$ is bounded and continuous for each $a$, we have
\[
\lim_{c\to 0}m_c(a)=m(a) \text{ for all }a\in\mathcal{A}.
\]
Moreover, our assumption that $L$ is Lipschitz in $a$ implies that each $m_c$ is Lipschitz with constant $\lambda$, since
\[
|m_c(a)-m_c(a')|
\le \int |L(a,\theta)-L(a',\theta)|d\pi_c(\theta|P)
\le \lambda\cdot d(a,a').
\]
The same bound holds for $m$. Since $\mathcal{A}$ is compact, pointwise convergence together with this common Lipschitz bound implies that
\[
\lim_{c\to 0}\Vert m_c-m\Vert_\infty=0.
\]
The result is then immediate from the argmax continuous mapping theorem (Theorem 3.2.2 of \citealt{van_der_vaart_weak_1996}). $\Box$

\paragraph{Proof of Proposition \ref{Proposition: Posterior Non-Concentration}}
Fix $P$ and write
\[
J=J_W(P),\qquad Q(\theta)=Q_W(\theta;P),\qquad g_\theta=g(\theta;P).
\]
Define
\[
I_c(P)=\int \pi_\theta(\vartheta)\pi_{\eta,c}(g(\vartheta;P)\mid \vartheta)\,d\vartheta
\]
and
\[
I^*(P)=\int \pi_\theta(\vartheta)\pi_\eta^*(g(\vartheta;P)\mid \vartheta)\,d\vartheta.
\]
Also let
\[
\pi_c(\theta\mid P)
=
\frac{\pi_\theta(\theta)\pi_{\eta,c}(g_\theta\mid \theta)}{I_c(P)}
\]
and
\[
\pi^*(\theta\mid P)
=
\frac{\pi_\theta(\theta)\pi_\eta^*(g_\theta\mid \theta)}{I^*(P)}.
\]
Since $\pi_\eta^*(\cdot\mid\theta)$ has strictly positive density and the posterior based on $\pi_\eta^*$ is well-defined, $I^*(P)\in(0,\infty)$. Finally, define
\[
w_{1,c}
=
\frac{(1-\phi)I_c(P)}{(1-\phi)I_c(P)+\phi I^*(P)}.
\]
Then
\[
\pi_c^\phi(\theta\mid P)
=
w_{1,c}\pi_c(\theta\mid P)+(1-w_{1,c})\pi^*(\theta\mid P).
\]

We first consider the case $J=J_W(P)>0$. Let
\[
Z_c(P)=\int f\left(\frac{1}{c}\eta'W(P)\eta\right)d\eta.
\]
Then
\[
\pi_{\eta,c}(g_\theta\mid \theta)
=
\frac{f\left(\frac{1}{c}Q(\theta)\right)}{Z_c(P)}.
\]
Making the change of variables $\eta=\sqrt{c}\,W(P)^{-1/2}u$ yields
\[
Z_c(P)
=
c^{k/2}|W(P)|^{-1/2}\int f(u'u)\,du
=C_Wc^{k/2},
\]
where $C_W\in(0,\infty)$ does not depend on $c$.

Because $Q(\theta)\ge J>0$ for all $\theta$ and $f$ is non-increasing,
\[
0\le \pi_{\eta,c}(g_\theta\mid \theta)
\le \frac{f(J/c)}{Z_c(P)}
=
C_W^{-1}c^{-k/2}f(J/c).
\]
Assumption \ref{Assumption: Thin-Tailed Priors} implies that
\[
u^m f(u)\to 0 \qquad\text{for every }m>0.
\]
Indeed, fix $m>0$ and $a>1$. Since $f(au)/f(u)\to 0$, for all sufficiently large $u$ we have
$f(au)\le a^{-(m+1)}f(u)$. Iterating this bound on the geometric grid $\{a^n u:n\ge 0\}$ and using that $f$ is non-increasing gives $x^m f(x)\to 0$ as $x\to\infty$. Taking $m=k/2$ and $u=J/c$ therefore yields
\[
c^{-k/2}f(J/c)\to 0.
\]
Hence
\[
\sup_\theta \pi_{\eta,c}(g_\theta\mid \theta)\to 0,
\]
so
\[
0\le I_c(P)
\le \sup_\vartheta \pi_{\eta,c}(g(\vartheta;P)\mid \vartheta)\int \pi_\theta(\vartheta)\,d\vartheta
\to 0.
\]
Therefore
\[
(1-\phi)I_c(P)+\phi I^*(P)\to \phi I^*(P)>0.
\]
For each fixed $\theta$,
\[
\pi_\theta(\theta)\left((1-\phi)\pi_{\eta,c}(g_\theta\mid \theta)+\phi\pi_\eta^*(g_\theta\mid \theta)\right)
\to
\phi\pi_\theta(\theta)\pi_\eta^*(g_\theta\mid \theta).
\]
Dividing numerator and denominator gives
\[
\lim_{c\to 0}\pi_c^\phi(\theta\mid P)
=
\frac{\pi_\theta(\theta)\pi_\eta^*(g(\theta;P)\mid \theta)}
{\int \pi_\theta(\vartheta)\pi_\eta^*(g(\vartheta;P)\mid \vartheta)\,d\vartheta},
\]
which proves the first claim.

Now we prove the result for $J=J_W(P)=0$.
Define
\[
I_c(P)=\int \pi_{\eta,c}\left(g(\theta ; P)\mid \theta\right)\pi_\theta(\theta)d\theta
\]
and
\[
I^*(P)=\int \pi_\eta^*\left(g(\theta ; P)\mid \theta\right)\pi_\theta(\theta)d\theta.
\]
Also let
\[
\pi^*(\theta \mid P)
=
\frac{\pi_\theta(\theta)\pi_\eta^*\left(g(\theta ; P)\mid \theta\right)}
{\int \pi_\theta(\tilde{\theta})\pi_\eta^*\left(g(\tilde{\theta} ; P)\mid \tilde{\theta}\right)d\tilde{\theta}}.
\]
Then we can write the posterior as
\[
\pi_c^\phi(\theta \mid P)
=
w_{1,c}\pi_c(\theta \mid P) + (1-w_{1,c})\pi^*(\theta \mid P),
\]
where $w_{1,c}
=
\frac{(1-\phi)I_c(P)}{(1-\phi)I_c(P)+\phi I^*(P)}.$ By Proposition \ref{Proposition: Posterior Concentration}, the posterior $\pi_c(\theta\mid P)$ concentrates on $\theta_W(P)$ as $c\to 0.$  Thus it suffices to show that $w_{1,c}\to 1.$

Since $J_W(P)=0$,
$Q_W(\theta ; P)
=
\|\theta-\theta_W(P)\|_{\tilde{W}(P)}^2
$
for $\tilde{W}(P)=X(P)'W(P)X(P).$  Let
$
Z_c(P)=\int f\left(\frac{1}{c}\eta'W(P)\eta\right)d\eta.
$
Then
\[
I_c(P)
=
\frac{\int \pi_\theta(\theta)f\left(\frac{1}{c}\|\theta-\theta_W(P)\|_{\tilde{W}(P)}^2\right)d\theta}{Z_c(P)}.
\]
Making the change of variables $\eta=\sqrt{c}W(P)^{-1/2}u$ in $Z_c(P)$ yields
\[
Z_c(P)=c^{k/2}\left|W(P)\right|^{-1/2}\int f(u'u)du.
\]

Next, since $\pi_\theta$ is continuous and $\pi_\theta(\theta_W(P))>0$, there exist $r>0$ and $m>0$ such that
$\pi_\theta(\theta)\ge m$
for all $\theta \in \tilde{B}_r(\theta_W(P))$, where
$\tilde{B}_r(\theta_W(P))=\left\{\theta:\|\theta-\theta_W(P)\|_{\tilde{W}(P)}<r\right\}.$
Hence
\begin{align*}
&\int \pi_\theta(\theta)f\left(\frac{1}{c}\|\theta-\theta_W(P)\|_{\tilde{W}(P)}^2\right)d\theta\\
&\qquad\ge
m\int 1\{\theta \in \tilde{B}_r(\theta_W(P))\}
f\left(\frac{1}{c}\|\theta-\theta_W(P)\|_{\tilde{W}(P)}^2\right)d\theta\\
&\qquad=
m c^{p/2}\int 1\{\|u\|_{\tilde{W}(P)}<r/\sqrt{c}\}f\left(\|u\|_{\tilde{W}(P)}^2\right)du.
\end{align*}
The last integral converges, by monotone convergence, to $\int f\left(\|u\|_{\tilde{W}(P)}^2\right)du>0.$
Thus there exists a constant $C_1>0$ such that, for all sufficiently small $c$,
\[
\int \pi_\theta(\theta)f\left(\frac{1}{c}\|\theta-\theta_W(P)\|_{\tilde{W}(P)}^2\right)d\theta
\ge C_1 c^{p/2}.
\]
Therefore, for some constant $C_2>0$, 
$I_c(P)\ge C_2 c^{(p-k)/2}.$
Since $k>p$, it follows that $I_c(P)\to \infty$, and hence $w_{1,c}\to 1$.

Finally, for any $\varepsilon>0$,
\begin{align*}
\int 1\{\theta \notin B_\varepsilon(\theta_W(P))\}d\pi_c^\phi(\theta \mid P)
&=
w_{1,c}\int 1\{\theta \notin B_\varepsilon(\theta_W(P))\}d\pi_c(\theta \mid P)\\
&\qquad +(1-w_{1,c})\int 1\{\theta \notin B_\varepsilon(\theta_W(P))\}d\pi^*(\theta \mid P).
\end{align*}
The first term converges to zero by Proposition \ref{Proposition: Posterior Concentration}, while the second converges to zero because $1-w_{1,c}\to 0$.  This proves the claim. $\Box$
 
\paragraph{Proof of Proposition \ref{Proposition: Posterior Credible Sets with RII Prior}}
Classic results in statistics, e.g. \cite{muirhead_aspects_1982}, Chapter 1.5, imply the result for the case of a single regressor.  For completeness, we prove the result for the general case.
The confidence interval (\ref{eq: Rotation-invariance interval}) corresponds to the set of values for $v'\theta$ where the test statistic
\begin{equation}\label{eq: test stat}
\frac{v'\theta_W(P)-v'\theta}{\sqrt{\frac{J_W(P)}{k-p}\sigma_v^2(P)}}
\end{equation}
has absolute value less than $t^*_{k-p,1-\frac{\beta}{2}}$.  Hence, if we can show that (\ref{eq: test stat}) follows a $t_{k-p}$ distribution under all priors $\pi$ satisfying Assumptions \ref{Assumption: MD posterior} and \ref{Assumption: mean zero}, the result is immediate.

Note that (\ref{eq: test stat}) is equal to the t-statistic from regression 
(\ref{eq: regression version}) in the text,
\begin{equation}\label{eq: test stat, alt}
\frac{v'\theta_W(P)-v'\theta}{\sqrt{\frac{J_W(P)}{k-p}\sigma_v^2(P)}}=\frac{v'(\tilde X(P)' \tilde X(P))^{-1}\tilde X(P)' \tilde\eta}{\sqrt{\frac{\tilde{\eta}'(I-M(P))\tilde{\eta}}{k-p} \sigma_v^2(P)}}
\end{equation}
where, again, $M(P)=\tilde X(P)(\tilde X(P)'\tilde X(P))^{-1}\tilde X(P)'$. Specifically, $\frac{\tilde{\eta}'(I-M(P))\tilde{\eta}}{k-p}$ corresponds to the unbiased variance estimate, so the denominator $\sqrt{\frac{\tilde{\eta}'(I-M(P))\tilde{\eta}}{k-p} \sigma_v^2(P)}$ corresponds to the homoskedastic standard error.  Note, moreover, that the t-statistic is scale-invariant in the error, so (\ref{eq: test stat, alt}) is equal to
\[
\frac{v'(\tilde X(P)' \tilde X(P))^{-1}\tilde X(P)' \frac{\tilde\eta}{\|\tilde\eta\|}}{\sqrt{\frac{\frac{\tilde\eta}{\|\tilde\eta\|}'(I-M(P))\frac{\tilde\eta}{\|\tilde\eta\|}}{k-p} \sigma_v^2(P)}}.
\]
Lemma \ref{Lemma: RII Prior} implies that $\frac{\tilde\eta}{\|\tilde\eta\|}$ is uniformly distributed on the unit sphere under any prior $\pi$ satisfying Assumptions \ref{Assumption: MD posterior} and \ref{Assumption: mean zero}. However, this is exactly the distribution of $\frac{Z}{\|Z\|}$ for $Z\sim N(0,I).$ It follows that under $\pi,$ (\ref{eq: test stat, alt}) has the same distribution as
\[
\frac{v'(\tilde X(P)' \tilde X(P))^{-1}\tilde X(P)' \frac{ Z}{\| Z\|}}{\sqrt{\frac{\frac{ Z}{\| Z\|}'(I-M(P))\frac{ Z}{\| Z\|}}{k-p} \sigma_v^2(P)}}=
\frac{v'(\tilde X(P)' \tilde X(P))^{-1}\tilde X(P)' Z}{\sqrt{\frac{Z'(I-M(P))Z}{k-p} \sigma_v^2(P)}},
\]
where we have again used scale invariance of the t-statistic.  However, the last expression is the t-statistic from the regression 
\[
\tilde{Y}(P) = \tilde{X}(P)\theta + Z,
\]
which is well-known to be $t_{k-p}$ distributed, completing the proof. $\Box$

\paragraph{Proof of Proposition \ref{Proposition: Posterior Credible Sets with RII Prior finite}}
The proof of Proposition \ref{Proposition: Posterior Credible Sets with RII Prior finite} proceeds identically to the proof of Proposition \ref{Proposition: Posterior Credible Sets with RII Prior}, noting that the structure of the supporting assumptions and claim is identical up to a change in notation and interpretation. $\Box$

\paragraph{Proof of Proposition \ref{Proposition: Posterior Credible Sets with RII Prior local}}
The proof of Proposition \ref{Proposition: Posterior Credible Sets with RII Prior local} proceeds identically to the proof of Proposition \ref{Proposition: Posterior Credible Sets with RII Prior}, noting again that the structure of the supporting assumptions and claim is identical up to a change in notation and interpretation, and setting $v' = K$.
$\Box$

\end{document}